\documentclass[journal=pasa,manuscript=research-article,year=2025,volume=XX]{cup-journal}

\usepackage{amsmath}
\usepackage{amssymb}
\usepackage{gensymb}
\usepackage[nopatch]{microtype}
\usepackage{booktabs}

\usepackage{verbatim}

\usepackage[hidelinks]{hyperref}
\hypersetup{colorlinks=true,linkcolor=red,citecolor=orange,filecolor=cyan,urlcolor=magenta}
\usepackage{orcidlink}

\title{Possible detection of HFQPOs associated with ‘unknown’ variability class of GRS 1915+105}

\author{Seshadri Majumder\orcidlink{0000-0002-2214-3593}}
\affiliation{Indian Institute of Technology Guwahati, Guwahati, 781039, India.}
\email[Seshadri Majumder]{smajumder@iitg.ac.in}

\author{Santabrata Das\orcidlink{0000-0003-4399-5047}}
\affiliation{Indian Institute of Technology Guwahati, Guwahati, 781039, India.}

\author{Anuj Nandi\orcidlink{0000-0002-3130-8986}}
\affiliation{Space Astronomy Group, ISITE Campus, U. R. Rao Satellite Centre, Outer Ring Road, Marathahalli, Bangalore, 560037, India.}

\keywords{accretion, accretion disks -- black hole physics -- X-rays: binaries -- stars: individual (GRS 1915+105)} 

\begin{document}

\begin{abstract}
We present a comprehensive spectro-temporal analysis of GRS $1915+105$ observed with {\it AstroSat} during June, $2017$. A detailed study of the temporal properties reveals the appearance of an { `unknown' variability class ($\tau$)} during $\rho \rightarrow \kappa$ class transition of the source. This new { `unknown' class ($\tau$)} is characterized by the irregular repetition of low count `dips' along with the adjacent `flare' like features in between two successive steady count rate durations, resulting in uniform `$C$' shaped distribution in the color-color diagram. { A detailed comparative study of the variability properties between the $\tau$ class and other known variability classes of GRS $1915+105$ indicates it as a distinct variability class of the source.} Further, we find evidence of the presence of possible HFQPO features at $\sim 71$ Hz with quality factor $\sim 13$, rms amplitude $\sim 4.69\%$, and significance  $3\sigma$, respectively. In addition, a harmonic-like feature at $\sim 152$ Hz is also seen with quality factor $\sim 21$, rms amplitude $\sim 5.75\%$ and significance $\sim 4.7\sigma$. The energy-dependent power spectral study reveals that the fundamental HFQPO and its harmonic are present in $3-15$ keV and $3-6$ keV energy ranges, respectively. Moreover, the wide-band ($0.7-50$ keV) spectral modelling comprising of thermal Comptonization component indicates the presence of a cool ($kT_{\rm e}\sim 1.7$ keV) and optically thick (optical depth $\sim 14$) Comptonizing `corona', which seems to be responsible in regulating the HFQPO features in GRS $1915$+$105$. Finally, we find the bolometric luminosity ($L_{\rm bol}$) to be about $42\% L_{\rm Edd}$ within $1-100$ keV, indicating the sub-Eddington accretion regime of the source.
\end{abstract}

\section{Introduction} \label{sec:intro}

GRS 1915$+$105, being an enigmatic Galactic black hole X-ray binary source (BH-XRB), exhibits exceptionally complex variability properties \cite[]{Greiner-etal1996, Paul-etal1997, Taam-etal1997, Paul-etal1998a, Paul-etal1998c, Paul-etal1998b, Yadav-etal1999, Morgan-etal1999, Muno-etal1999, Belloni-etal2000, Naik-etal2002, Chakrabarti-etal2004, Rodriguez-etal2004, Belloni-etal2013a, Weng-etal2018, Zhang-etal2022, Athulya-etal2022, Majumder-etal2022, Athulya-etal2023} in different time scales ranging from seconds to hours. The rich variability properties of GRS 1915$+$105 render this source a unique astrophysical laboratory to study the accretion dynamics around black holes. The distinct variability pattern observed in the light curves and the corresponding Color-Color Diagrams (CCDs) manifest $15$ unique classes \citep{Belloni-etal2000,Klein-wolt-etal2002, Hannikainen-etal2005, Pahari-etal2009} during {\it RXTE} era. So far, eight confirmed variability classes, namely $\theta$, $\chi$, $\omega$, $\delta$, $\beta$, $\rho$, $\kappa$ and $\gamma$ of GRS $1915$+$105$, are observed with {\it AstroSat} during $2016-2022$ \cite[]{Athulya-etal2022, Majumder-etal2022, Athulya-etal2023}. Interestingly, an intermediate variability state, characterized by unstructured broad patterns in the light curve are also observed with {\it AstroSat} during the transition from quiet $\chi$ to structured $\rho$ class transition \citep{Rawat-etal2019, Athulya-etal2022}.

\begin{table*}
	\centering
	\caption{Observation details of GRS $1915+105$ observed by {\it AstroSat} and {\it NuSTAR} during June, $2017$. Column $1$, $2$, $4$, $5$ and $6$ are for ObsID, Epoch, Orbit number, MJD and exposure time. In column $3$, `Seg.' denotes the light curve segments for Epochs AS$1$ and AS$2$. The mean detected ($r_{\rm det}$) and incident ($r_{\rm in}$) count rates, along with the mean hardness ratios (HRs) over the entire exposures for each variability class, are tabulated. HR1\_var, HR2\_var and $\rm r_{det}$\_var are the fractional variabilities of the respective quantities. In column $14-15$, the variability classes and the presence of Low$/$High-frequency QPO features in respective observations are given.	See the text for details.}
	\renewcommand{\arraystretch}{1.05}
	\resizebox{1.0\textwidth}{!}{ 
    
	\begin{tabular}{l @{\hspace{0.2cm}} c @{\hspace{0.3cm}} c @{\hspace{0.2cm}} c c @{\hspace{0.2cm}} c @{\hspace{0.2cm}} c @{\hspace{0.2cm}} c @{\hspace{0.2cm}} c @{\hspace{0.2cm}} c @{\hspace{0.2cm}} c @{\hspace{0.2cm}} c @{\hspace{0.2cm}} c @{\hspace{0.2cm}} c @{\hspace{0.2cm}} c @{\hspace{0.2cm}} c}
		\toprule
		ObsID & Epoch & Seg. & Orbit & MJD & Effective & $\rm r_{det}$ & $\rm r_{in}$ & HR1 & HR2 & { HR1\_var} & { HR2\_var} & { $\rm r_{det}$\_var} & Class & LF$/$HF \\
		& & & & (Start) & Exposure (ks)& (cts/s)& (cts/s)& (B/A)$^*$& (C/A)$^*$ & ($\%$) & ($\%$) & ($\%$) & & QPO\\
		\midrule

            10408-01-30-00 & RX$0$ & $2$ & $-$ & $50313.31$ & $3.4$ & $16919$ & $-$ & $0.68$ & $0.09$ & $2$ & $7$ & $7$ & $\chi_{3}$ & LF\\

             20402-01-33-00 & RX$1$ & $2$ & $-$ & $50617.61$ & $3.3$ & $14676$ & $-$ & $1.14$ & $0.11$ & $24$ & $27$ & $60$ & $\kappa$ & LF \\

             20402-01-53-01 & RX$2$ & $1$ & $-$ & $50757.21$ & $3.2$ & $21288$ & $-$ & $1.02$ & $0.08$ & $17$ & $21$ & $27$ & $\mu$ & $-$ \\
        
		 G07\_046T01\_9000001236 & AS$0$ & $-$ & $8876$ & $57892.74$ & $3.63$ & $1292$ & $1328$ & $0.61$ & $0.09$ & $13$ & $28$ & $42$ & $\rho$ & LF \\
         
		 G07\_028T01\_9000001272 & AS$1$ & $1a$ & $9127$ & $57909.77$ & $2.43$ & $2156$ & $2259$ & $0.62$ & $0.04$ & $16$ & $37$ & $60$ & unknown & $-$ \\
         
		 & & $1b$  & $9127-9128$ & $57909.83$ & $3.36$ & $2236$ & $2348$ & $0.62$ & $0.04$ & $16$ & $44$ & $62$ & unknown & HF \\
         
		 & & $1c$  & $9130$ & $57909.89$ & $2.95$ & $2235$ & $2346$ & $0.62$ & $0.04$ & $17$ & $49$ & $72$ & unknown & $-$ \\
         
		 & & $1d$  &  $9130-9131$ & $57909.98$ & $0.94$ & $2133$ & $2235$ & $0.60$ & $0.04$ & $15$ & $30$ & $48$ & unknown & $-$ \\
         
	    G07\_046T01\_9000001274 & AS$2$ & $2a$ & $9131-9136$ & $57909.99$ & $6.53$ & $2135$ & $2236$ & $0.61$ & $0.04$ & $15$ & $29$ & $47$ & unknown & $-$ \\
		
		 & & $2b$ & $9136-9139$ & $57910.43$ & $9.03$ & $2167$ & $2272$ & $0.60$ & $0.04$ & $14$ & $31$ & $45$ & unknown & HF \\
		
		 30302018002 & NU$1$ & $-$ & $-$ & $57928.67$ & $18.4$ & $572$ & $-$ & $0.78$ & $0.02$ & $7$ & $31$ & $4$ & $\delta$ & $-$ \\
         
		 G07\_028T01\_9000001370 & AS$3$ & $-$ & $9629$ & $57943.69$ & $0.98$ & $2814$ & $2993$ & $0.76$ & $0.04$ & $21$ & $43$ & $47$ & $\kappa$ & $-$ \\
		
		\toprule
	\end{tabular}
    }
	
	\label{table:Obs_details}
	
	\begin{list}{}{}
		\item[$*$] A, B and C are the count rates in $3-6$ keV, $6-15$ keV and $15-60$ keV energy ranges, respectively.
	\end{list}
\end{table*}

Meanwhile, several attempts have been made to explain the rich variability properties and transition between different classes of GRS $1915+105$. For instance, \cite{Greiner-etal1996} argued for an accretion disc instability mechanism to be responsible for the quasi-periodic variation in the X-ray light curves of varying duration and repetition timescales. Further, a disc-corona configuration was attributed to the rapid regular/irregular variabilities of the source \citep{Taam-etal1997, Vilhu-etal1998}. Interestingly, \citealt[]{Nandi-etal2000} proposed that the variability classes of GRS $1915+105$ can be classified into four different types, which are closely connected to the `softness ratio' diagram associated with the Kepleran and sub-Keplerian accretion flows. In addition, \citealt[]{Chakrabarti_Nandi_2000} suggested that the transition between three fundamental accretion states of GRS $1915+105$ results in the observed variability patterns. Subsequently, the feedback of the failed disc winds to the accretion flow is conjectured to be associated with the well-known `On' state of the source, possibly responsible for the flickering in light curves of several classes \citep{Nandi-etal2001}. Further, \citealt[]{Janiuk-etal2002} argued that the lack of direct transition from the canonical `state $C$' to `state $B$' in the CCDs of GRS $1915+105$ can be explained considering the time evolution of a viscous disc driven by the instabilities at the innermost region. However, \citealt[]{Chakrabarti-etal2005} suggests that the transition between variability classes on a timescale much shorter than the viscous time is attributed to the change in the accretion rate of the sub-Keplerian flow. This further corroborates the presence of nearly freely falling low angular momentum accretion flow as the source of rapid variability in the count rates during the transitions. In addition, \cite{Neilsen-etal2011} proposed that the radiation pressure-driven mass loss from the inner disc at near Eddington luminosity is sufficient enough to explain the oscillation in accretion rate, resulting in the hard pulses seen in the $\rho$ class light curve of `heartbeat' state in GRS $1915+105$. However, detection of $\rho$ class with {\it AstroSat} in the sub-Eddington accretion state further challenges the proposed models to explain the complex nature of the variability classes \citep{Athulya-etal2022}.

\begin{figure*}
	\begin{center}
		\includegraphics[scale=0.5]{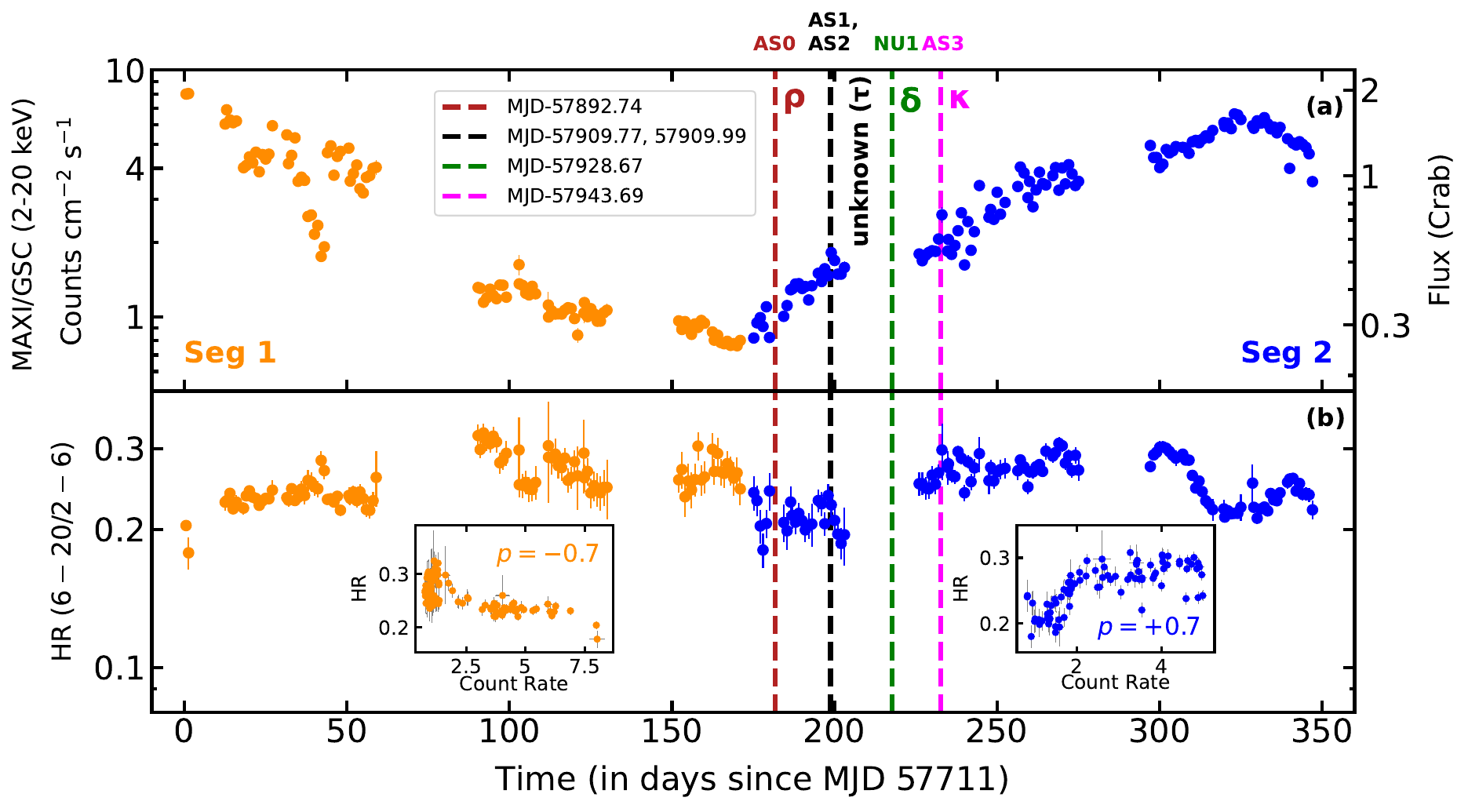}
	\end{center}
	\caption{\textbf{(a)} {\it MAXI/GSC} daily light curve in the energy band of $2-20$ keV in flux units of counts $\rm cm^{-2}$ $\rm s^{-1}$. \textbf{(b)} The variation of the hardness ratio defined as the ratio of $6-20$ keV to $2-6$ keV count rates with the time of observation. { In panel (b), the insets show the variation of the hardness ratio with the corresponding count rates, indicated by distinct colors (orange and blue). $p$ denotes the Pearson correlation coefficient between the count rate and hardness ratio distributions.} Different colored vertical dashed lines represent the Epochs of observations with {\it AstroSat} and {\it NuSTAR}, respectively. See the text for details.}
	\label{fig:maxi_lc}
\end{figure*}

GRS $1915$+$105$ occasionally exhibits the High-Frequency Quasi-periodic Oscillations (HFQPOs) feature, which is believed to be one of the most powerful diagnostic tools to probe the effect of strong gravity in the vicinity of black holes. Indeed, several other BH-XRBs, namely XTE J$1550-564$, GRO J$1655-40$, IGR J$17091-3624$ and H $1743-322$ display HFQPOs in the frequency range of $\sim 65-450$ Hz \cite[]{Remillard-etal_2006}. Among these sources, GRS $1915$+$105$ particularly draws considerable attention because of its stable HFQPOs at $\sim 70$ Hz, which is observed on multiple occasions by both {\it RXTE} \cite[and references therein]{Strohmayer2001a, Remillard-etal2002, Belloni-etal2013a} and {\it AstroSat} \cite[and references therein]{Sreehari-etal2020, Majumder-etal2022, Majumder-etal2024, Majumder-etal2025, Harikesh-etal2025}. Interestingly, some of the `softer' variability classes of GRS $1915$+$105$, such as $\delta$, $\kappa$, $\omega$, $\gamma$, $\mu$, $\rho$, and $\nu$, are found to exhibit the HFQPO features \citep{Belloni-etal2013a, Majumder-etal2022}, which are attributed to the modulation of inner `Comptonizing corona' \citep{Sreehari-etal2020, Majumder-etal2022}. Moreover, simultaneous detection of additional features at $\sim 34$ Hz and $41$ Hz along with the $70$ Hz HFQPO are also reported occasionally \citep{Strohmayer2001b, Belloni-etal2013b}. 

Furthermore, the phase/time lag properties associated with the HFQPOs of GRS $1915$+$105$ are also extensively studied with both {\it RXTE} and {\it AstroSat}. For example, hard phase lag is attributed to the HFQPO feature at $67$ Hz, whereas soft lag is seen at $34$ Hz with {\it RXTE} observations \citep{Mendez-etal2013}. Although, energy dependent soft time lags of maximum $\sim 3$ ms duration have been observed at HFQPO ($\sim 70$ Hz) in different variability classes ($\delta$, $\omega$, $\kappa$, $\gamma$) of GRS $1915$+$105$ with {\it AstroSat} \citep{Majumder-etal2024}. This is further elucidated by the reflection of hard photons in the cooler accretion disc, giving rise to the observed soft time delay \citep{Majumder-etal2024}.

Intriguingly, GRS $1915$+$105$ remained active \cite[]{Balakrishnan-etal2021, Motta-etal2021,Athulya-etal2022,Parrinello-etal2023} for last few decades since its discovery by {\it WATCH} onbroad {\it GRANAT} in $1992$ \cite[]{Castro-Tirado-etal1992} before its quiescence phase from $2018$, followed by sporadic re-brightenings in $2019$ \citep{Athulya-etal2023}. The source is continuously monitored by {\it AstroSat} from $2016$ onwards providing quality observations. Therefore, it becomes immensely appealing to study the spectro-temporal characteristics over different variability classes of the source. Indeed, a comprehensive wide-band spectro-temporal study is carried out by \citealt[]{Athulya-etal2022} to probe the evolution in the accretion dynamics of the source. Further, a multimission study of the `obscured' phase of GRS $1915+105$ results in the detection of several emission and absorption spectral lines along with mHz QPOs in the power spectra \citep{Athulya-etal2023}.

In this work, we present the results of an in-depth spectro-temporal study of two {\it AstroSat} observations of GRS $1915+105$ during June, $2017$. While doing so, we carry out a detailed temporal study and find the presence of an `unknown' variability class than the previously known $15$ classes of the source. In addition, we find the evidence of possible HFQPOs at $71$ Hz and its harmonic-like feature at $\sim 152$ Hz. Further, the energy dependent characteristics of the observed HFQPO signatures are also examined. Subsequently, we study the wide-band ($0.7 -50$ keV) spectral properties of the source and put efforts into explaining the anticipated origin of the possible HFQPO features observed in GRS $1915+105$ during the `unknown' variability class.

The paper is organized as follows. In \S2, we mention the details of the selected observations and describe the data reduction procedure. In \S3 and \S4, we present the in-depth temporal and spectral analysis with results, respectively. We discuss the obtained results in \S5, and finally, conclude in \S6.

\section{Observation and Data Reduction} \label{s:Data-reduction}
 
In the present work, we analyze two Guaranteed Time (GT) observations of GRS 1915$+$105 during June $2017$ with {\it SXT} and {\it LAXPC} onboard {\it AstroSat} \citep{Agrawal-etal2006, Singh-etal2014, Agrawal-etal2017}. Three more adjacent observations available with {\it AstroSat} and {\it NuSTAR} during May$-$July, $2017$ are also analyzed. Additionally, we analyze three {\it RXTE/PCA} observations (RX$0-$RX$2$) in our study, which are available in the \texttt{HEASARC} archive\footnote{\url{https://heasarc.gsfc.nasa.gov/db-perl/W3Browse/w3browse.pl}}. The details of all these observations are tabulated in Table \ref{table:Obs_details} along with Epoch numbers. The Epochs of these observations are marked using vertical lines in the MAXI/GSC light curve (Fig. \ref{fig:maxi_lc}).

We extract {\it LAXPC} \citep{Antia-etal2017} level-1 data in event analysis mode, available in {\it AstroSat} public archive\footnote{\url{https://webapps.issdc.gov.in/astro_archive/archive/Home.jsp}} using the standard data reduction software \texttt{LaxpcSoftv3.4.4}\footnote{\url{http://www.tifr.res.in/~astrosat_laxpc/LaxpcSoft.html}}. The details of {\it LAXPC} data extraction procedure and analysis methods are mentioned in \cite{Sreehari-etal2019, Sreehari-etal2020, Majumder-etal2022}. It is worth mentioning that, due to the constraints of Earth occultation and passage of the satellite in the region of South Atlantic Anomaly (SAA), continuous observation is not possible with {\it AstroSat}. Subsequently, the data corresponding to a given observation of {\it AstroSat/LAXPC} are generally divided into multiple Good Time Interval (GTI) segments of the observation time using the analysis software. For example, the entire data of AS1 observation contains $4$ GTI segments ($1a$, $1b$, $1c$ and $1d$) (see Table \ref{table:Obs_details}), which are distributed over $4$ orbits of observation. Further, we mention that the timing analyses including the study of variability properties and power density spectra are carried out for such individual segments excluding the data gaps, whereas the spectral analysis is performed considering the entire observation data of AS1 and AS2.

We utilize both {\it LAXPC10} and {\it LAXPC20} data for the timing analyses, whereas only {\it LAXPC20} data are used in the spectral study following \cite{Antia-etal21, Majumder-etal2022}. Note that, the data from the top layer of the detectors and that corresponds to the single events are considered for the analysis. Accordingly, we generate the source and background spectra along with instrument response files for {\it LAXPC20} following \cite{Antia-etal2017}. We analyze the {\it SXT} \citep{Singh-etal2016, Singh-etal2017} data following the guidelines provided by the {\it SXT} instrument team\footnote{\url{https://www.tifr.res.in/~astrosat_sxt/index.html}}. The {\it SXT} light curves, spectra and ancillary response files are extracted using \texttt{XSELECT V2.5b} from a $12$ arcmin circular source region using the level-2 cleaned event files available at ISSDC\footnote{\url{https://webapps.issdc.gov.in/astro_archive/archive/Home.jsp.}} data archive.

The {\it NuSTAR} data analysis is carried out using the latest mission specific software available at \texttt{HEASOFT V6.32.1}\footnote{\url{https://heasarc.gsfc.nasa.gov/docs/software/heasoft/}}. The task {\it nupipeline} is used to generate the cleaned event file from the data of both {\it FPMA} and {\it FPMB} instruments of {\it NuSTAR}. Following \cite{Nathan-etal2022}, a circular region of $50$ arcsec at the source position and away from the source is considered to obtain the source and background products, respectively. We use the {\it nuproduct} task to generate the source and background light curves. 

{\it RXTE/PCA} data available both in Binned and Event mode are analyzed for timing studies. We extract the light curves of different time resolutions across various energy bands using the \texttt{saextrct} and \texttt{seextrct} tasks available in \texttt{FTOOLS} for the Binned mode and Event mode data, respectively (see \citealt[]{Nandi-etal2012} for details). Note that during the observation of $\rm RX0-RX2$ Epochs, all five PCUs of {\it RXTE/PCA} were on.

\begin{figure}
	\begin{center}
		\includegraphics[scale=0.47]{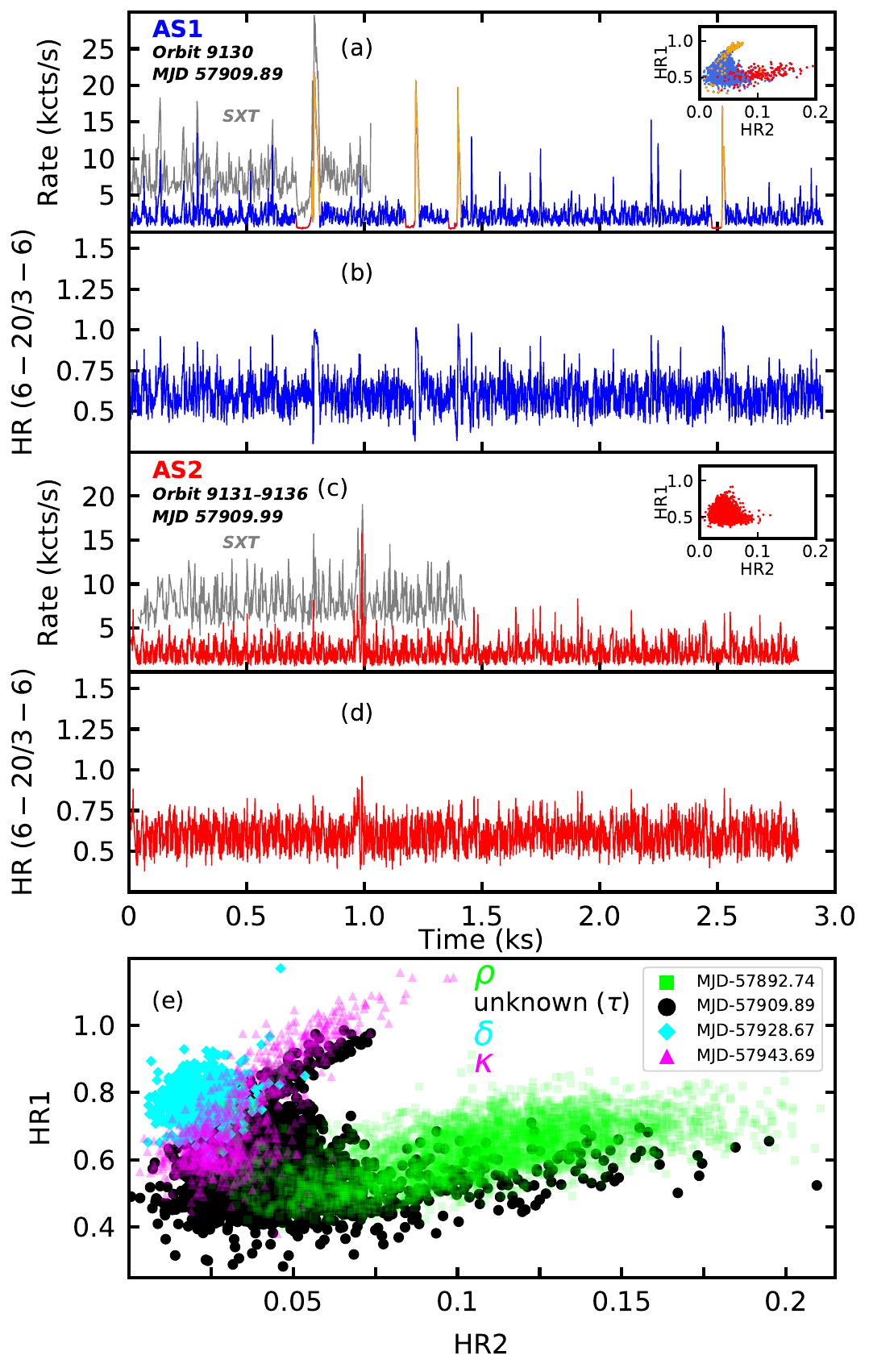}
	\end{center}
	\caption{Background subtracted and dead-time corrected $1$ s binned light curves ({\it LAXPC} in $3-60$ keV) of GRS 1915$+$105 observed with {\it AstroSat} during Epoch AS$1$ (a) and AS$2$ (c), respectively. The {\it SXT} light curves of AS$1$ and AS$2$, simultaneous with {\it LAXPC} is shown with gray color. The CCD of the same observations are shown at the top right insets, respectively. The variation of the corresponding hardness ratios are shown in panel (b) and (d). The comparison of CCDs in different variability classes under consideration is shown in panel (e). See text for details.}
	\label{fig:lcurvCCD}
\end{figure}

\section{{ Temporal Analysis and Results}} \label{s:timing}

\subsection{MAXI/GSC Monitoring of GRS 1915+105} \label{s:maxi}

We study the evolution of the source GRS $1915+105$ with daily MAXI/GSC monitoring. In Fig. \ref{fig:maxi_lc}a, we present the light curve in $2-20$ keV energy range during MJD $57711-58061$. In Fig. \ref{fig:maxi_lc}b, we show the variation of hardness ratio (HR) defined as the ratio of counts in $6-20$ keV and $2-6$ keV energy bands. { The dashed vertical lines in brown, black, green and magenta denote the Epochs corresponding to {\it AstroSat} and {\it NuSTAR} observations under consideration. Further, we carry out a correlation study between HR and count rates for two distinct regions of the MAXI/GSC light curves, represented by orange (Seg 1) and blue (Seg 2) colors in Fig. \ref{fig:maxi_lc}. The variation of HR against the count rates for each of these two segments is shown in the insets of Fig. \ref{fig:maxi_lc}b. While computing the Pearson correlation coefficient ($p$) of these distributions, we consider all the data points in Seg 1, whereas count rates exceeding $5$ counts $\rm cm^{-2}$ $\rm s^{-1}$ are excluded for Seg 2. With this, we find a strong negative (positive) correlation between HR and count rate for Seg 1 (Seg 2) with $p = -0.7\;(+0.7)$ for one-tailed chance probability of $3.1 \times 10^{-16}$ ($5.1 \times 10^{-15}$). 

Moreover, we note that the {\it AstroSat} observations considered in this study fall within the Seg 2 region.  This suggests that, during this period, the source was exhibiting `softer' spectral characteristics with a $\sim 2.5$ times increase in the count rate and a marginal variation in the HR. The detection of `softer' variability classes (`unknown', $\delta$, and $\kappa$) during this period further supports the softening behavior of the source. At the initial phase of Seg 2 the MAXI/GSC count rate was $\lesssim 1$ counts $\rm cm^{-2}$ $\rm s^{-1}$, during which $\rho$ class variability characterized by quasi-periodic flares with relatively harder characteristics in the CCD (see \citealt[]{Belloni-etal2000}) was detected. It is important to note that the canonical $\chi$ class variability of harder spectral characteristics in GRS $1915+105$ is generally associated with the less variable quiescent state C, often persisting over longer timescales ranging from hours to months \citep{Belloni-etal2000}. In contrast, the $\phi$ class, which resembles a typical soft state variability pattern, is seen to be connected with the outbursting state A of the source \citep{Belloni-etal2000}. Notably, the `softer' variability classes (`unknown', $\delta$, and $\kappa$) remain in between these two and correspond to rapid transitions between the outbursting states and the short-lived quiescent state C on much smaller times scales of few tens of seconds.}

\subsection{Temporal Variability}

We generate $1~{\rm s}$ binned background subtracted {\it LAXPC10} and {\it LAXPC20} combined light curves in $3-60$ keV energy range to investigate the variability properties of the source. The dead-time corrected average incident and detected count rates \cite[and references therein]{Sreehari-etal2020} estimated in $3-60$ keV energy range are given in Table \ref{table:Obs_details}. We generate the CCDs of each observation by defining the soft and hard colors as HR1$=B/A$ and HR2$=C/A$, where $A$, $B$ and $C$ are the count rates in $3 - 6$ keV, $6 - 15$ keV and $15 - 60$ keV energy bands, respectively \citep{Sreehari-etal2019, Sreehari-etal2020}. Similarly, the CCD of {\it NuSTAR} observation is calculated considering same energy bands. The average { values} of $HR1$ and $HR2$ for each observation are listed in Table \ref{table:Obs_details}. { Further, we compute the fractional variability amplitude in detected count rate and HR values following \cite{Vaughan-etal2003} and tabulate in Table \ref{table:Obs_details}.} We present the light curves ({\it SXT} in grey, and {\it LAXPC} in blue and red) of AS$1$ and AS$2$ Epochs in Fig. \ref{fig:lcurvCCD}a and Fig. \ref{fig:lcurvCCD}c along with the CCDs at top right insets. The variation of hardness ratio (HR$=D/A$) corresponding to AS$1$ and AS$2$ is depicted in Fig. \ref{fig:lcurvCCD}b and Fig. \ref{fig:lcurvCCD}d, respectively, where $D$ being the count rates in $6-20$ keV energy band. 

\begin{figure}
	\begin{center}
		\includegraphics[scale=0.47]{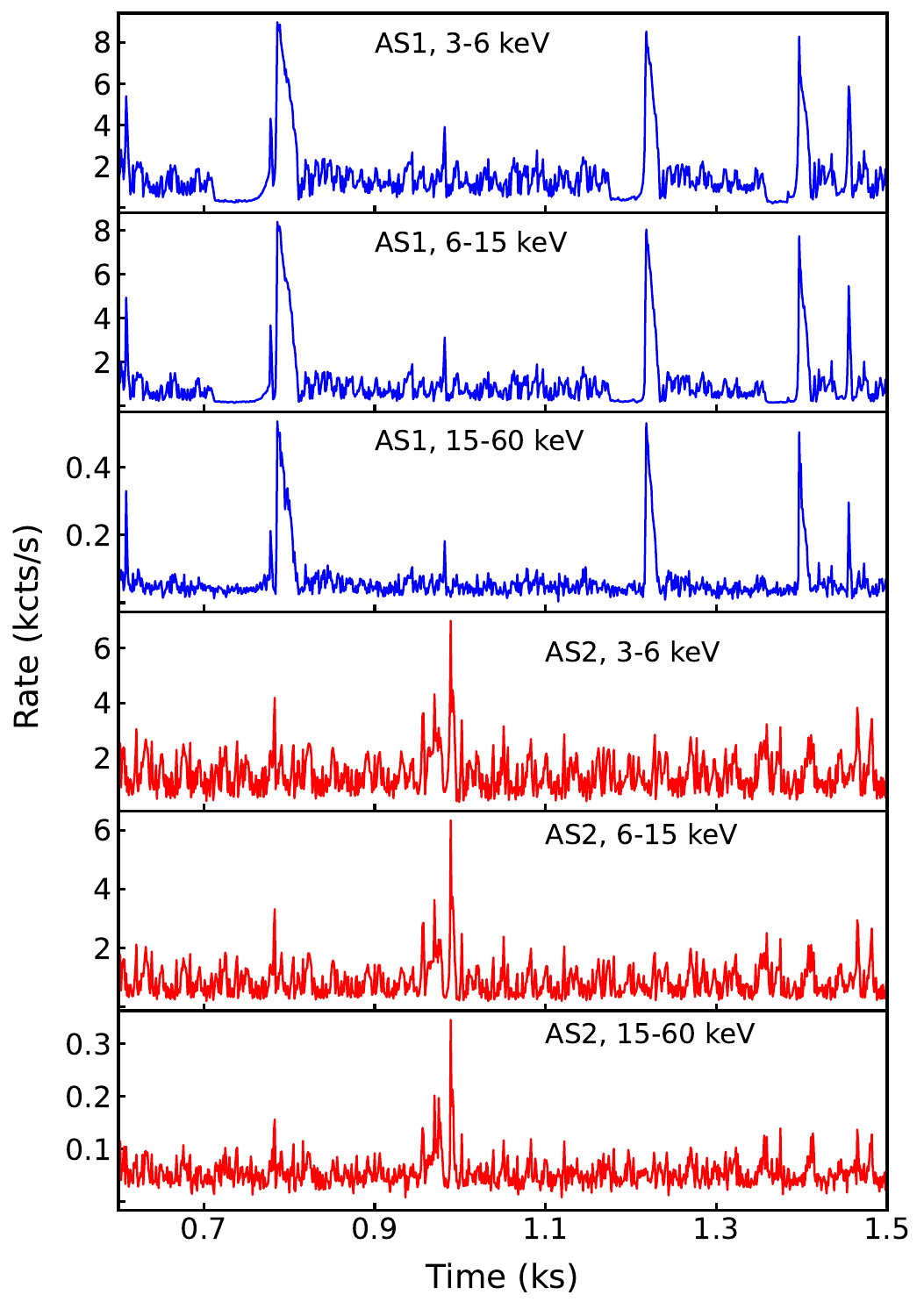}
	\end{center}
	\caption{{ Light curves from the AS1 and AS2 observations in the $3-6$ keV, $6-15$ keV, and $15-60$ keV energy ranges are shown in the top, middle, and bottom panels, respectively. These segments correspond to the $0.6-1.5$ ks interval of the full light curves shown in Fig. \ref{fig:lcurvCCD}, highlighting the structured variability. See text for details.}}
	\label{fig:lcurv_ene}
\end{figure}
Investigation of the variability patterns observed in the {\it LAXPC} light curve of AS$1$ reveals the appearance of several irregular `dips' (low counts) of maximum $\sim 60~{\rm s}$ duration (Fig. \ref{fig:lcurvCCD}a) between two consecutive `non-dip' (steady counts) segments. In addition, a `flare' like feature is also seen immediately after the `dip'. However, Epoch AS$2$ exhibits almost steady light curves (Fig. \ref{fig:lcurvCCD}c) similar to the `non-dip' duration of AS$1$. The patten of the CCD of AS$2$ (Fig. \ref{fig:lcurvCCD}c) resembles with that of AS$1$ (`C' shape pattern, see Fig. \ref{fig:lcurvCCD}a) without its two elongated branches. { Further, we investigated the variability properties of AS1 and AS2 observation across different energy bands. For this, in Fig. \ref{fig:lcurv_ene}, we present the light curves of AS1 and AS2 observations in $3-6$ keV, $6-15$ keV, and $15-60$ keV energy bands, focusing on the time interval ($0.6-1.5$ ks) during which the repetitive dip and spike-like features are observed (see Fig. \ref{fig:lcurvCCD}). Interestingly, we find that the overall structured variability pattern remains broadly similar across all energy bands, with only the count rate varying between them. More specifically, we observe that the low-count `dip' features in the light curve are most prominent and pronounced in the $3-6$ keV and $6-15$ keV energy bands, but they disappear in the $15-60$ keV band. In contrast, the characteristics of the spike-like features remain similar across all energy bands. In addition, during the AS2 observation, the structured variability shows marginal differences in different energy bands.}

The observed variability pattern seems to be different from the known $15$ variability classes of GRS 1915$+$105. Indeed, the overall shape of the CCD in AS$1$ is coarsely similar to the $\lambda$ class, however, the variability pattern and the duration of `dip' and `non-dip' segments are distinctly different from the $\lambda$ and $\kappa$ classes. Further, the `flare' like features observed just after the `dips' in AS$1$ seem to be similar to $\rho$ class \citep{Belloni-etal2000, Athulya-etal2022}. In a preceding observation with {\it AstroSat} (AS$0$ in Fig. \ref{fig:maxi_lc}, see Table \ref{table:Obs_details}) $16$ days before AS$1$, the source was in $\rho$ class \cite[]{Athulya-etal2022}. In addition, the source was observed in $\kappa$ class (Epoch AS$3$ in Fig. \ref{fig:maxi_lc}, see Table \ref{table:Obs_details}) after $33$ days of AS$2$. Interestingly, the {\it NuSTAR} observation (NU$1$ in Fig. \ref{fig:maxi_lc}, see Table \ref{table:Obs_details}) which is $15$ days prior to AS$3$ shows $\delta$ class variability. { All the above findings clearly indicate that the source exhibits multiple class transitions within a months timescale during Epoch AS$0$ to AS$3$. Therefore, we infer that possibly GRS $1915+105$ went through an `unknown' variability class rather than the previously known $15$ distinct classes. However, we also note that due to the lack of continuous observations of the source, we are unable to track all possible class transitions that might have occurred on timescales of hours to days, as is typically seen for GRS 1915$+$105 \citep{Chakrabarti-etal2005}.}

\begin{figure}
	\begin{center}
		\includegraphics[scale=0.45]{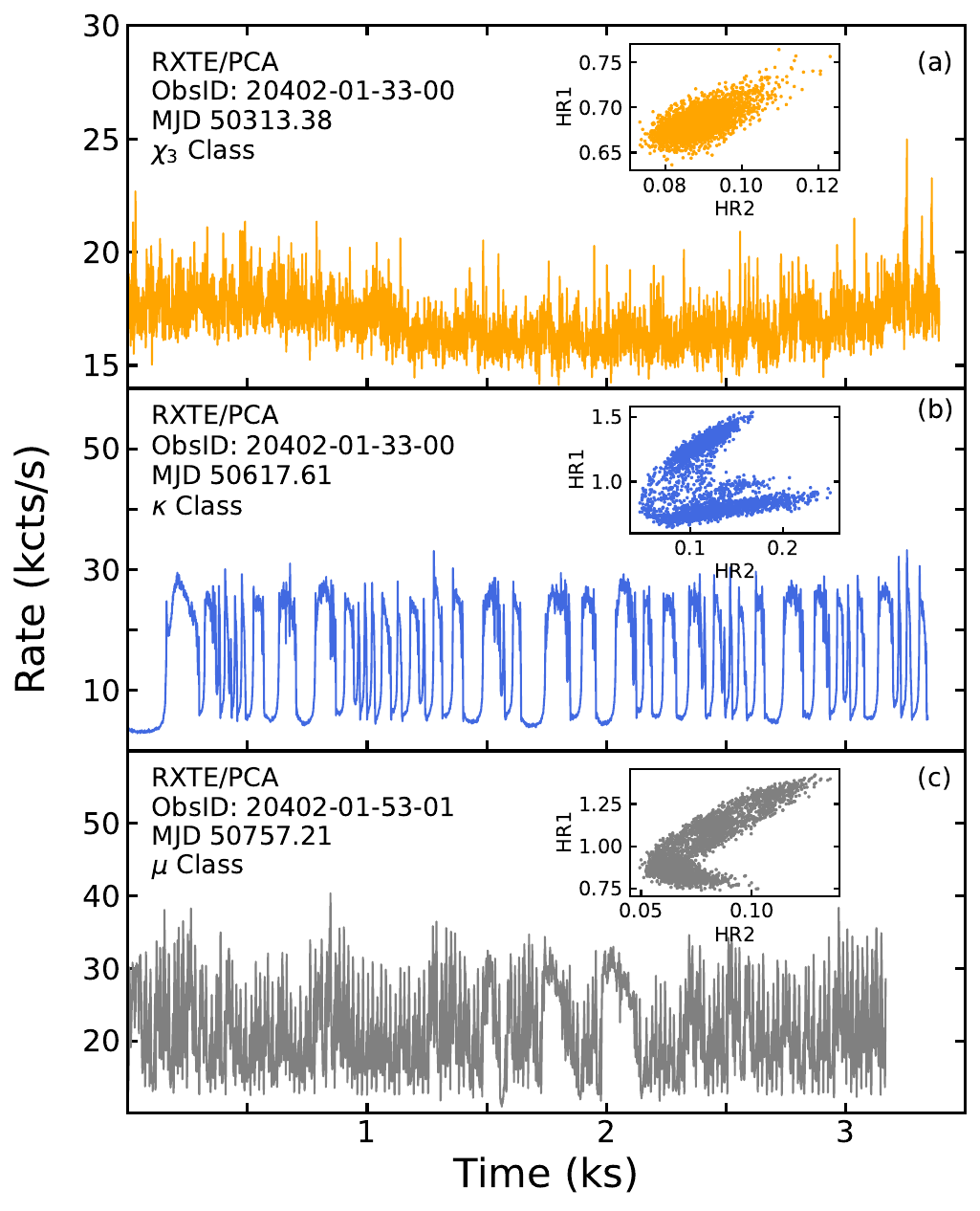}
	\end{center}
	\caption{{\it RXTE/PCA} light curves of $1$ s time bin of the source GRS $1915+105$ in $3-60$ keV energy range corresponding to $\chi_{3}$, $\kappa$ and $\mu$ variability classes. The CCDs of the respective classes are presented in the insets of each panel. See text for details.}
	\label{fig:rxte_lc}
\end{figure}

In Fig. \ref{fig:lcurvCCD}e, we compare the CCDs generated from Epochs AS$0$, AS$1$, NU$1$ and AS$3$ (Table \ref{table:Obs_details}). It is evident that the source underwent transition from $\rho$ to $\kappa$ class via an `unknown' and $\delta$ class variabilities. A detailed investigation of the CCDs indicates that the `harder' tail (black filled circles with HR$1 \lesssim 0.6$ and HR$2 \gtrsim 0.07$) and the `softer' branch (HR$1 \gtrsim 0.6$ and HR$2 \lesssim 0.07$) of AS$1$ overlap with the distributions of a $\rho$ (filled green squares) and $\kappa$ (filled magenta triangles) classes, respectively. We observe that $\delta$ class (filled cyan diamonds), generally `softer' in nature, appears in between AS$0$ and AS$3$, and it makes the `unknown' class more complex as the CCD shows the mixed properties of both $\rho$ and $\kappa$ classes. This further corroborates the appearance of an `unknown' variability class which is characterized by the irregular soft `dips' along with `flare' like spikes in the light curve between two consecutive steady count rates.

\begin{figure*}
		\includegraphics[scale=0.4]{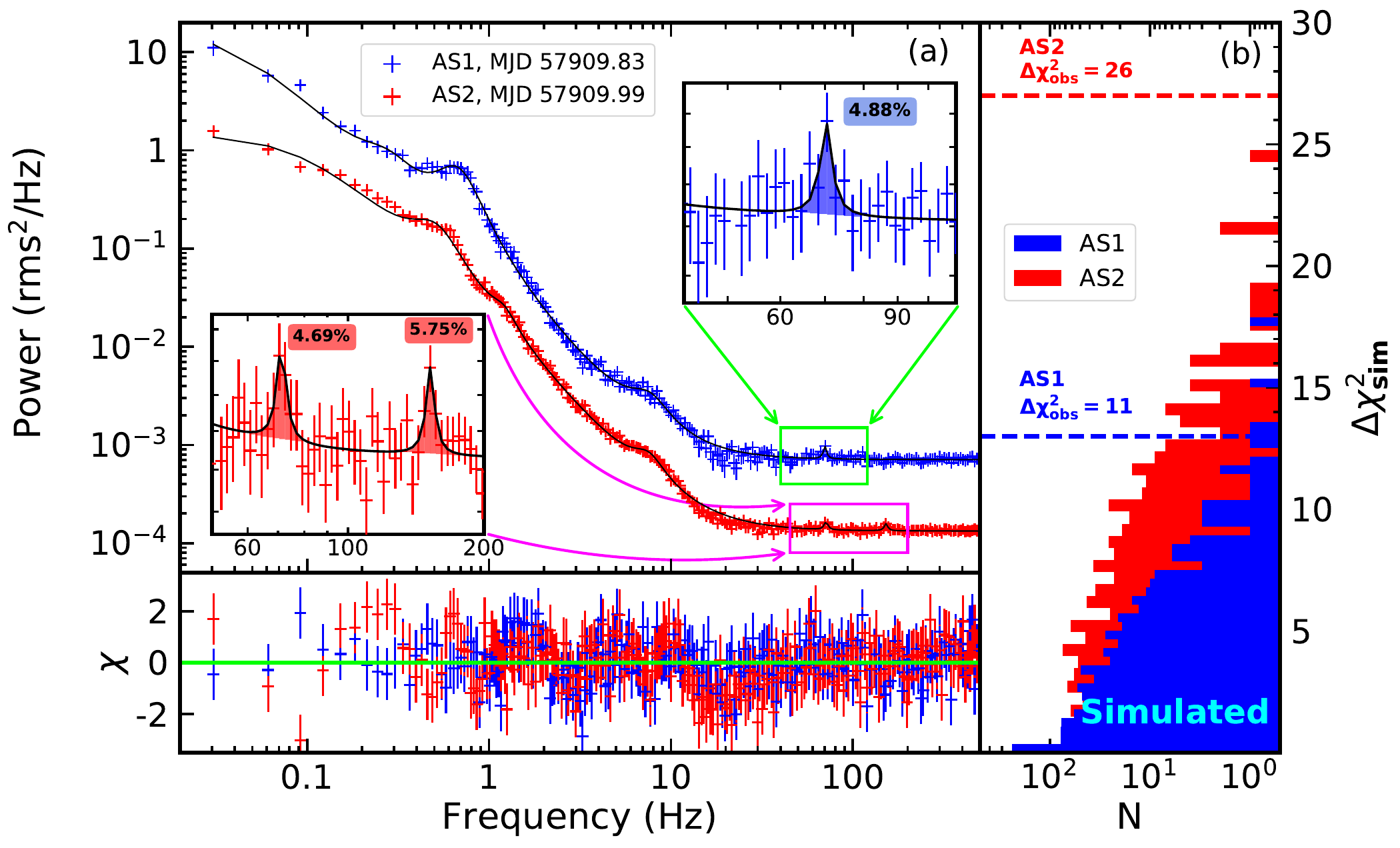}
	\caption{{\it Panel (a)}: Power density spectra of Epoch AS$1$ and AS$2$ in the broad-band frequency range ($0.01-500$ Hz). Each PDS is obtained in $3-60$ keV energy band using {\it LAXPC10} and {\it LAXPC20} combined observation. Zoomed view of the detected HFQPO and/or harmonic features are shown in the inset. For clarity purpose, PDS of Epoch AS$1$ is re-scaled by multiplying factor $5$. {\it Panel (b)}: The distribution of $\Delta \chi_{\rm sim}^2$ with the number of occurrences (N), obtained from $1000$ simulated power spectra. The horizontal dash lines denote the $\Delta \chi_{\rm obs}^2$ values obtained using observational date for AS$1$ (blue) and AS$2$ (red), respectively. See text for details.}
	\label{fig:PDS_QPO}
\end{figure*}

\subsection{Comparison with { $\chi_{3}$, $\kappa$ and $\mu$} classes}

In this section, we perform a comparative study between the variability properties of the `unknown' class (AS1 and AS2) and the canonical $\chi_{3}$, $\kappa$ and $\mu$ classes observed with {\it RXTE/PCA} (see Table \ref{table:Obs_details}). In doing so, we present $1$ s time binned light curves of the $\chi_{3}$, $\kappa$, and $\mu$ classes in Fig. \ref{fig:rxte_lc} with CCDs at the insets, obtained from the {\it RXTE/PCA} data in $3-60$ keV energy band. Note that, to compare with AS1 and AS2 observations, we obtain the CCDs of $\chi_{3}$, $\kappa$ and $\mu$ classes using the definition of soft color and hard color as $HR1=B/A$ and $HR2=C/A$, respectively, where $A$, $B$ and $C$ represent the light curves in $3-6$ keV, $6-15$ keV and $15-60$ keV energy bands, respectively. { It is important to note that the observed structured variability pattern in the light curve and the shape of the CCD for a given class largely depend on both effective area of the instruments used and, to some extent, the overall flux level of the source. For instance, the {\it AstroSat/LAXPC} observations were conducted at a relatively lower source intensity of $\sim 555$ mCrab and with a much larger effective area beyond $\sim 10$ keV \citep{Beri-etal2019}. In contrast, the {\it RXTE/PCA} observations were performed during a much brighter phase of the source ($\sim 1.2$ Crab). Because of that, despite the sharply declining effective area of {\it RXTE/PCA} beyond $\sim 10$ keV, the count rates at higher energies may increase significantly, thereby altering the hardness ratios. Keeping these limitations in mind, in this section, we focus solely on a qualitative comparison between the CCDs of different classes observed with {\it RXTE/PCA} and the `unknown' variability class observed with {\it AstroSat/LAXPC}.}

In Fig. \ref{fig:lcurvCCD} and Fig. \ref{fig:rxte_lc}, we observe that the structured variability in the `unknown' class remains different from the individual variability patterns of $\chi_{3}$ and $\mu$ classes mainly due to the appearance of irregular small duration dips followed by high count spikes in AS1. { Interestingly, these dip features appear somewhat similar to the low-count dips of varying durations observed in the $\kappa$ class (see Fig.~\ref{fig:rxte_lc}). However, we find that the typical duration of these dips is about $\sim 60$ s in AS1, whereas it varies between $\sim 50-90$ s in the $\kappa$ class. Moreover, the fractional noise amplitudes (defined as the standard deviation normalized by the mean count rate) of these dip intervals are found to be $\sim 9.3\%$ and $\sim 15.2-17.4\%$ for the `unknown' and $\kappa$ variability classes observed with {\it AstroSat} and {\it RXTE}, respectively. This suggests that although an apparent similarity exists between the dip segments of the canonical $\kappa$ class and those of the `unknown' variability class observed during AS1, the intrinsic noise amplitude is nearly twice as high in the case of the $\kappa$ class, indicating different underlying variability processes. Moreover, we also find that the noise amplitudes of the dip segments in AS1 are nearly twice that of the corresponding Poisson noise in these intervals, implying the presence of intrinsic variability during the dip durations beyond random statistical fluctuations. Given the similarity between the small-amplitude spikes observed during the steady interval of the `unknown' class (AS2) and those seen in the $\chi_3$ variability class, we compare the Poisson noise-subtracted intrinsic fractional variability amplitude ($F_{\rm var}$) for these two cases. For this, we compute $F_{\rm var}$ following \cite{Vaughan-etal2003} for both $\chi_3$ class and AS2 observation. We find that $F_{\rm var} \sim 7\%$ for $\chi_3$ class, whereas it increases significantly to $\sim 45-47\%$ during the AS2 observation. This indicates that the small-amplitude spikes and fluctuations observed in the steady interval of the `unknown' variability class are intrinsically different from the variability characteristics of the $\chi_3$ class. 

Further, we find that the distribution of $HR1$ and $HR2$ in the CCD (see Fig. \ref{fig:lcurvCCD}c) of AS1 observation remains completely different from the canonical $\chi_{3}$ class but resemble with that of $\kappa$ and $\mu$ class to some extent (see Fig. \ref{fig:rxte_lc}). However, the upper and lower branches of the $\mu$ class CCD remain asymmetric, whereas a more extended lower branch towards higher $HR2$ values is observed for the AS1 observation.}

{ Furthermore, we compare the characteristics of the present `unknown' class in context of the canonical states of GRS 1915$+$105 proposed by \cite{Belloni-etal2000}. We note that, \cite{Belloni-etal2000} identified three basic spectral states while classifying the variability properties of GRS 1915$+$105. Among these, state C and state B correspond to the quiescent and outburst states, respectively. More specifically, state C is associated with low flux and is harder in nature, typically located towards the higher HR2 region of the CCD. In contrast, state B is characterized by high flux, higher HR1, and lower HR2 values. In addition to these, another state known as state A is observed during transitions between states B and C. In general, state A exhibits low flux but softer spectral colors and is typically located in the lower-left region of the CCD. 

It is important to mention that the $\lambda$ and $\kappa$, basic classes of GRS $1915+105$ \citep{Belloni-etal2000}, exhibit low-flux dips associated with the quiescent state C and repeated high-count episodes corresponding to the outburst state B \citep{Belloni-etal2000}. Typically, the low-count dips of varying durations remain located along the lower branch of the CCD in these two classes. Similarly, we observe that the low flux dips of the `unknown' class during AS1 form the harder branch of the CCD (see inset in Fig. \ref{fig:lcurvCCD}a) and resemble the quiescent state C of GRS $1915+105$, as proposed by \cite{Belloni-etal2000}. On the other hand, high count spikes generally reside in the higher HR1 region of the CCD, indicating the possible presence of the outburst state B. However, we find the presence of a cluster of points in the CCD corresponding to steady durations characterized by small-amplitude spikes in the light curve of AS1 (see inset in Fig. \ref{fig:lcurvCCD}a). This region of the CCD, which exhibits softer spectral characteristics and low flux, broadly resembles state A, albeit with a much greater spread. Typically, state A appears for short durations in the $\lambda$ and $\kappa$ classes and therefore contributes only marginally to the CCD \citep{Belloni-etal2000}. Nevertheless, in the present case, the irregular and structured variability observed in AS1 and AS2 observation makes it more challenging to confirm the presence of state A.}

\subsection{Power Density Spectra} \label{s:pds}

We generate $1$ ms resolution light curves with the combined data from both {\it LAXPC10} and {\it LAXPC20} in $3-60$ keV energy range. This allows us to examine the power spectral properties in the wide-band frequency range of $0.01-500$ Hz. Following \cite{Sreehari-etal2020, Majumder-etal2022}, we obtain the power density spectra (PDS) for all segments of Epoch AS$1$ and AS$2$. In particular, we consider $32768$ bins per interval for generating individual PDS that are again averaged out to obtain the resultant PDS of each segment of observations. Note that, for a continuous data of exposure $T_{exp}$ and time resolution $\Delta t$, the total number of individual PDS available for averaging becomes $T_{exp}/32768 \Delta t$. Thus, approximately $104$ and $274$ number of PDS are averaged out to obtain the resultant PDS for AS1($1b$) and AS2($2b$) observations, respectively. While doing this, if a light curve duration is found to be longer than a multiple of $32768$ bins, the excess data towards the end is truncated. This procedure is followed to adopt the computational efficiency of the Fast Fourier Transform (FFT) algorithm and to maintain consistency across different intervals, producing individual PDS of uniform frequency grids \citep{VanderKlis1989, Uttley-etal2014}. Finally, a geometric binning factor of $1.03$ is chosen to rebin the PDS in the frequency space. We mention that the errors in power are estimated from the standard deviation of power values across independent frequency bins. Further, we note that because of the geometrical rebinning, the power values within a bin are averaged and accordingly, the propagated errors are expected to introduce correlations between adjacent bins of overlapping frequency intervals. Note that, the dead-time correction of the light curves is implemented during the PDS generation following \cite{Agrawal-etal2018, Sreehari-etal2019, Sreehari-etal2020}. Accordingly, dead-time affected Poisson noise level is subtracted from the PDS in Leahy normalization \citep{Leahy-etal1983} and the corresponding effects on rms amplitude are also corrected following \cite{Zhang-etal1995, Sreehari-etal2019, Sreehari-etal2020} to obtain the PDS in the rms space.

\begin{table*}
	\begin{center}	
		\caption{Details of the detected HFQPO characteristics along with the fit statistics for Epoch AS$1$ and AS$2$. Here, $\nu_{\rm HFQPO}$, FWHM, $norm$, $Q$-$factor$, Sig. and $\rm HFQPO_{\rm rms}$ denote the frequency, width, normalization, quality factor, significance and rms amplitude of the detected HFQPO and/or harmonic features. $\rm Total_{\rm rms}$ denotes the rms amplitude of the entire PDS. $\chi^2/$dof indicates the reduced chi-square ($\chi^2_{\rm red}$) of the best fitted PDS. P$_\texttt{simftest}$ represents the \texttt{simftest} probability. All the errors are computed with $68$\% confidence level. See text for details.}
		
		\renewcommand{\arraystretch}{1.4}
		
		\begin{tabular}{l @{\hspace{0.3cm}} c @{\hspace{0.4cm}} c @{\hspace{0.4cm}} c @{\hspace{0.3cm}} c @{\hspace{0.4cm}} c @{\hspace{0.4cm}} c @{\hspace{0.4cm}} c @{\hspace{0.4cm}} c @{\hspace{0.4cm}} c}
			
			\toprule
			
			Epoch &  $\chi^2/$d.o.f & P$_\texttt{simftest}$ & $\nu_{\rm HFQPO}$ & $\rm FWHM$ ($\Delta$) & $norm$ ($K$) & $Q$-$factor$ & Sig. & $\rm HFQPO_{\rm rms}$ & $\rm Total_{\rm rms}$ \\
			
			&  & ($\times 10^{-3}$) & (Hz) & (Hz) & ($\times10^{-4}$) &  & ($\sigma$) & ($\%$) & ($\%$) \\
			
			\midrule 
			
			AS$1$ ($1b$) & $166/228$ & $16$ &  $70.24_{-0.85}^{+0.87}$ & $4.01_{-1.21}^{+1.18}$ & $3.04^{+1.56}_{-1.39}$ & $17.52$ & $2.2$ & $4.88\pm0.97$ & $56.33\pm10.48$ \\
			
			AS$2$ ($2b$) & $220/224$ & $1$ &  $71.15_{-1.05}^{+1.01}$ &  $5.33_{-2.85}^{+3.32}$ & $2.34_{-0.79}^{+1.05}$ & $13.35$ & $3$ &  $4.69\pm0.78$ & $57.78\pm7.06$ \\
			
			& &  & $152.30_{-2.06}^{+1.55}$ & $7.23_{-3.88}^{+4.25}$ & $2.59^{+0.73}_{-0.55}$ & $21.07$ & $4.7$ & $5.75\pm0.81$ & $57.78\pm7.06$ \\
			
			\toprule
			
		\end{tabular}
		
		\label{specpar}
		\label{table:PDS_parameters}
		
	\end{center}
	
\end{table*}

Each PDS in rms space is modeled in \texttt{XSPEC V12.13.1} using a \texttt{constant} and \texttt{Lorentzian} functions. The \texttt{Lorentzian} function inside \texttt{XSPEC} reads as,
\begin{equation}
L(\nu)= \frac{K}{2\pi} \frac{\Delta}{(\nu -\nu_{\rm c})^2+ (\Delta/2)^2} ,
\end{equation}
where, $\nu$ is the frequency and $\nu_{\rm c}$ represents the centroid of the \texttt{Lorentzian} profile. Here, $\Delta$ is the full width at half maximum (FWHM) and $K$ is the associated normalization. We find that multiple \texttt{Lorentzian} components at several frequencies are required to fit the PDS continuum over the entire wide-band frequency range. In particular, it is observed that one zero-centroid \texttt{Lorentzian} component ($L_0$) along with additional \texttt{Lorentzian} components at several distinct frequencies are required to model the red noise continuum. For example, two \texttt{Lorentzian} functions $L_1$ and $L_2$ are required to fit the broad bump-like features present at $\sim 0.65$ Hz and $\sim 7.35$ Hz in the power spectra of AS1 ($1b$) observation. In addition, one additional \texttt{Lorentzian} component ($L_3$) is needed to fit the power spectral break present at $\sim 0.25$ Hz. It may be noted that the inclusion of each of these additional components ($L_1$-$L_3$) significantly improves the overall fit with a reduction in the fitted $\chi^{2}$ by $\sim 29$ ($L_1$), $157$ ($L_2$) and $28$ ($L_3$) for degrees of freedom reduced by $3$ corresponding to each component. Interestingly, the significance of these broad features are found to be $12.6\sigma$ ($L_1$), $10.4\sigma$ ($L_2$) and $3.2\sigma$ ($L_3$), whereas the quality factor remains $< 2$, indicating broad characteristics. Note that, we adopt the standard definition of quality factor ($Q=\nu_{\rm c}/\Delta$) and significance ($\sigma=K/err_{\rm neg}$, where $err_{\rm neg}$ being the negative error in normalization) in the estimation of the respective quantities following \cite{Casella-etal2005, Belloni-etal2013a, Sreehari-etal2020, Majumder-etal2022}. Similarly, two broad ($Q \lesssim 2$) but significant ($\sim 9\sigma$) features at $\sim 0.48$ Hz and $\sim 7.35$ Hz are present for AS2 ($2b$). However, unlike AS1 ($1b$), the power spectral break frequency is found to be around $\sim 1.2$ Hz in AS2 ($2b$). As before, modeling of these individual features renders a reduction of $\chi^{2}$ within $50-275$ for $3$ degrees of freedom. Moreover, the above findings suggest that these low-frequency broad bump-like features are perhaps connected to inherent variability properties of the `unknown' class and distinguish it from the other known classes of GRS $1915+105$. 

Interestingly, we observe that significant residuals are still present near $\sim 70$ Hz in both observations (segments $1b$ and $2b$, see Table \ref{table:Obs_details}) which are further modeled using a  \texttt{Lorentzian} function. The inclusion of additional \texttt{Lorentzian} results in a reduction in chi-square ($\chi^2$) of $\sim 11$ (for AS$1$) with the degrees of freedom ($dof$) reduced by $3$. Further, we notice the presence of an additional peak at $\sim 150$ Hz along with $\sim 70$ Hz during the Epoch AS$2$ (segment $2b$). As before, simultaneous fitting of both these two peaks reduces $\chi^2$ by $\sim 26$ in AS$2$. Finally, the best fit of the entire PDS is obtained with $\chi^2/dof$ of $166/228$ (segment $1b$ of AS$1$) and $220/224$ (segment $2b$ of AS$2$). In Fig. \ref{fig:PDS_QPO}a, we present the best-fitted PDS of Epoch AS$1$ and AS$2$, respectively. It is observed that the overall shape of the PDS remain similar for both observations including the broad features present around $\sim 0.6$ Hz and { $\sim 7$ Hz}, respectively. The residual variations corresponding to the model-fitted PDS are shown in the bottom panel of Fig. \ref{fig:PDS_QPO}a.

The power spectral modeling indicates the evidence of a HFQPO feature at $70.24_{-0.85}^{+0.87}$ Hz and $71.15_{-1.05}^{+1.01}$ Hz having significance (quality factor) of $2.2\sigma$ ($17.52$) and $3\sigma$ ($13.35$) in the best fitted PDS of Epoch AS$1$ ($1b$) and AS$2$ ($2b$), respectively. In addition, we compute the rms amplitude by taking the square root of the integrated power of the \texttt{Lorentzian} function describing the HFQPO feature and find it to be $(4.88\pm0.97)\%$ (AS$1$) and $(4.69\pm0.78)\%$ (AS$2$). Further, we also notice the presence of a possible harmonic-like feature at $152.30_{-2.06}^{+1.55}$ Hz (Fig. \ref{fig:PDS_QPO}) along with the fundamental peak at $\sim 70$ Hz during Epoch AS$2$. The significance (quality factor) and percentage rms amplitude of the harmonic-like feature are found to be $4.7\sigma$ ($21.07$) and $(5.75\pm0.81)\%$, respectively. In addition, the total rms amplitude over the entire PDS of frequency range $0.01-500$ Hz is found to be $(56.33\pm10.48)\%$ and $(57.78\pm7.06)\%$ during Epochs AS$1$ and AS$2$, respectively. The best-fitted and estimated HFQPO parameters and the fit statistics for both observations are tabulated in Table \ref{table:PDS_parameters}.

Next, we perform \texttt{simftest} \cite[]{Lotti-etal2016,Athulya-etal2022} available within the \texttt{XSPEC} environment to further investigate the possible detection of the HFQPO and harmonic features. In general, the \texttt{simftest} script\footnote{\url{https://heasarc.gsfc.nasa.gov/xanadu/xspec/manual/XspecManual.html}} simulates synthetic spectra by adding random Poisson noise to the best-fitted model. This approximates the real statistical fluctuations, expected in the observational data. Next, the fitting is performed for the individual simulated spectra, which allows us to determine the significance of a fitted model component. With this, we generate $1000$ simulated power spectra and obtain the simulated $\Delta \chi_{\rm sim}^2$ to compare with the observed $\Delta \chi_{\rm obs}^2$. Here, $\Delta \chi_{\rm sim}^2$ and $\Delta \chi_{\rm obs}^2$ are the change in chi-square statistics with and without fitting the HFQPO and/or harmonic-like feature in the simulated and observed PDS, respectively. The obtained results are depicted in Fig. \ref{fig:PDS_QPO}b using histogram. We find that for both AS$1$ and AS$2$, $\Delta \chi_{\rm obs}^2$ resides away from the main simulated distribution of $\Delta \chi_{\rm sim}^2$. These findings yield a `low' chance probability of $\sim 0.001$ and $0.016$ of getting the features by random statistical fluctuations in AS$1$ and AS$2$, respectively. Moreover, based on these statistical analyses and the best-fitted HFQPO parameters (see Table \ref{table:PDS_parameters}), we conclude the possible evidence of HFQPO and/or harmonic features in AS$1$ and AS$2$ observations, respectively. However, low significance of the observed features (see Table \ref{table:PDS_parameters}) possibly implies the feeble nature of the HFQPOs and therefore the results should be interpreted accordingly. It is worth mentioning that the PDS for all the segments of AS$1$ and AS$2$ along with rest of the observations in Table \ref{table:Obs_details} are analyzed. However, we find the presence of HFQPOs for the segments $1b$ and $2b$ only, and therefore, the results are presented in Fig. \ref{fig:PDS_QPO} and Table \ref{table:PDS_parameters} for these observation segments.
\begin{figure}
	\begin{center}
            \includegraphics[scale=0.75]{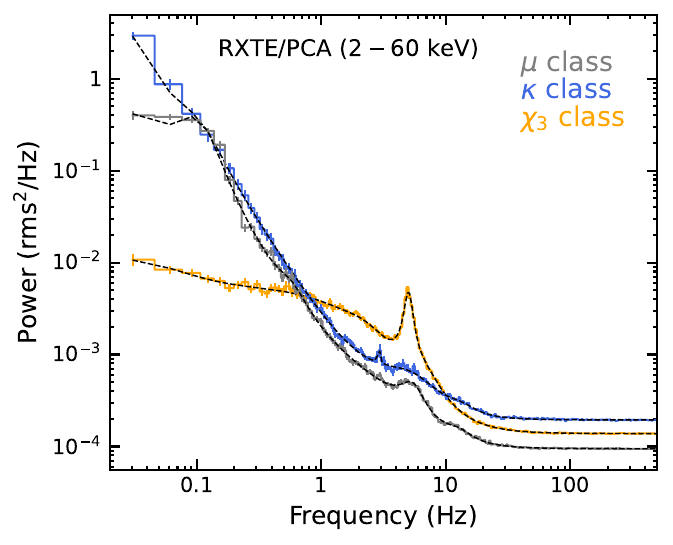}
	\end{center}
	\caption{The PDS in $2-60$ keV energy band of $\chi_{3}$, $\kappa$, and $\mu$ classes corresponding to the same observations for which the light curves are presented in Fig. \ref{fig:rxte_lc}. See text for details.}
	\label{fig:rxte_pds}
\end{figure}

Further, we examine the power spectral properties of $\chi_{3}$, $\kappa$ and $\mu$ classes using {\it RXTE/PCA} data and compare with that of the `unknown' variability class observed during AS1 and AS2. The PDS computation is performed following the similar methodology described above and modeled using the combination of \texttt{constant} and \texttt{Lorentzian} components. In Fig. \ref{fig:rxte_pds}, we present the PDS corresponding to $\chi_{3}$, $\kappa$ and $\mu$ classes in $2-60$ keV energy band. We observe that a strong low-frequency QPO is present at $\sim 5$ Hz in the PDS of $\chi_{3}$ class, whereas broad bump-like features are visible around $\sim 6$ Hz in the $\kappa$ and $\mu$ class observations. In addition, a QPO-like feature seems to be present at $\sim 3$ Hz in the PDS of $\kappa$ class. Moreover, it is evident that the overall shape of the PDS in $\mu$ and $\kappa$ class including the low-frequency power spectral break near $\sim 0.1$ Hz and $\sim 0.6$ Hz along with bump-like features around $\sim 6$ Hz remain marginal identical to the PDS obtained during AS1 and AS2 (see Fig. \ref{fig:PDS_QPO}) without the presence of HFQPOs. However, the appearance of a strong QPO feature at $\sim 5$ Hz along with a flat-topped noise distribution at lower frequencies makes the PDS of $\chi_{3}$ observation completely distinct from the rest of the classes.

\subsection{On the Modeling of LFQPO}

{ In this section, we examine the accuracy in modeling the low-frequency ($< 20$ Hz) domain of the PDS. In doing so, we consider the PDS of the AS1 observation up to $20$ Hz for modeling, which is shown in the top panel of Fig. \ref{fig:pds_compare}. The bottom panels of Fig. \ref{fig:pds_compare} present the variation of model fitted residuals with different model combinations as described below. First, we fit one simple \texttt{powerlaw} to model the red noise part of the PDS and obtain a reduced chi-square ($\chi_{\rm red}^{2}$) of $7$ with large variations in the residuals. Next, we include one \texttt{Lorentzian} (L1) component at $0.6$ Hz to fit the LFQPO-like feature. This yields a worst fit for a $\chi_{\rm red}^{2}=3.6$ with significant residual variation at $0.2-0.3$ Hz and $7-8$ Hz (see Fig. \ref{fig:pds_compare}). Hence, we include two additional \texttt{Lorentzian} components (L2 and L3) to model these bump-like features, resulting in a $\chi_{\rm red}^{2}=0.9$ (see Fig. \ref{fig:pds_compare}). 
\begin{figure}
	\begin{center}
            \includegraphics[scale=0.45]{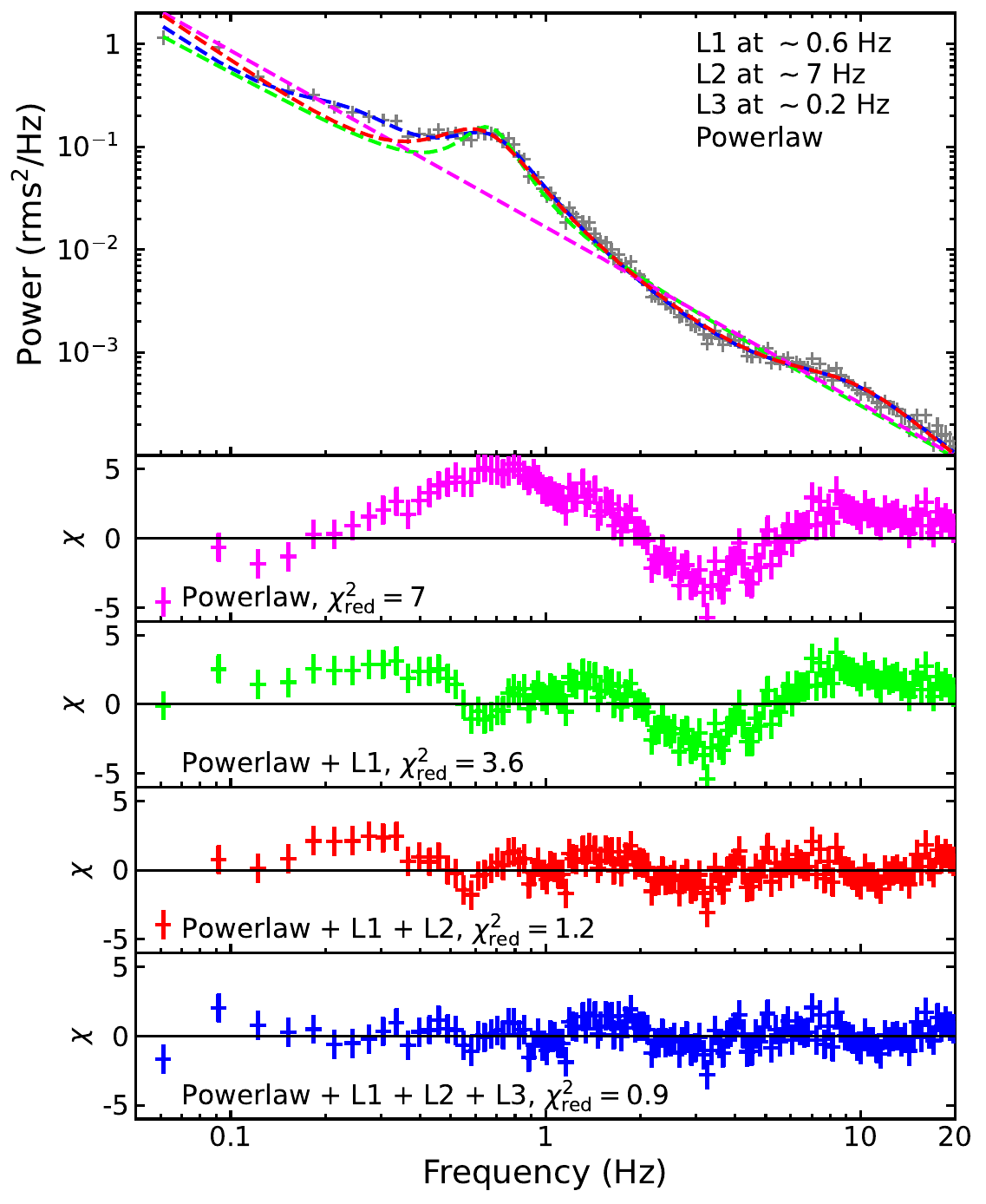}
	\end{center}
	\caption{{ \textit{Top panel:} PDS of the AS1 observation up to 20 Hz, highlighting the low-frequency variability properties. Dashed lines in different colors represent the model combinations used in the PDS modeling. \textit{Bottom panels:} Variation of the model-fitted residuals for different model combinations, as indicated in the figure. See text for details.}}
	\label{fig:pds_compare}
\end{figure}
Moreover, we note that excluding either the L2 or L3 component from the fit results in $\chi_{\rm red}^{2}$ value exceeding $1.3$. From this modeling, we find the significance of the broad LFQPO-like feature at $0.63$ Hz to be around $11.2\sigma$. Note that the broadband ($0.01-500$ Hz) PDS modeling (see \S3.4) resulted in the centroid frequency and significance of this feature as $0.65$ Hz and $12.6\sigma$, respectively. This suggests that the characteristic frequency of the LFQPO-like feature and its properties remain independent of the choice of modeling the red noise component. Furthermore, a careful investigation of the model-fitted PDS in Fig. \ref{fig:pds_compare} indicates that the \texttt{powerlaw} dominates in the modeling of the red noise component only below $\sim 0.18$ Hz. On the other hand, the presence of two broad bump-like features at $\sim 0.2$ and $\sim 7$ Hz dominates over \texttt{powerlaw} in modeling the overall red noise variation. This further suggests that the red noise in the PDS deviates significantly from a simple \texttt{powerlaw} form and needs to be modeled using multiple \texttt{Lorentzian} components at distinct frequencies, as also demonstrated in \S3.4 during the broadband PDS modeling.}

\subsection{Energy Dependent Power Spectra} \label{s:ene_pds}

\begin{figure}
	\begin{center}
		\includegraphics[scale=0.36]{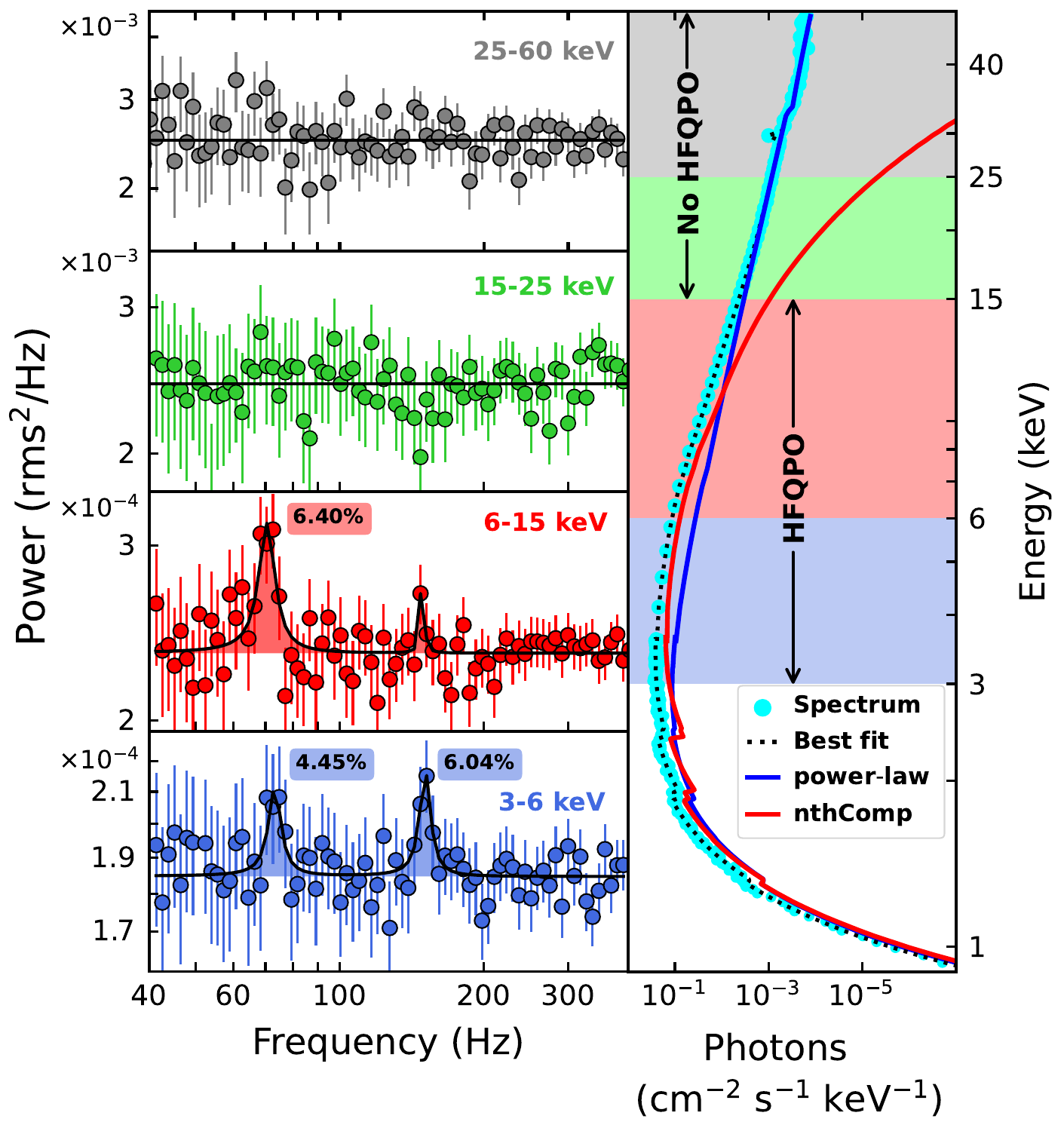}
	\end{center}	
	\caption{{\it Left panel}: Energy dependent power density spectra of Epoch AS$2$. In top to bottom panels, the best fitted PDS in different energy ranges are depicted. {\it Right panel}: Best fitted wide-band ($0.7-50$ keV) energy spectrum of combined {\it SXT} and {\it LAXPC} observations. The dotted (black) and solid curves (blue/red) represent the best fitted models and spectral components used in the modeling. The shaded regions with different colors represent the energy bands for which the PDS are computed. See text for details.
	}
	\label{fig:PDS_zoom}
\end{figure}

We intend to study the energy-dependent properties of the possible HFQPO and harmonic features obtained in the preceding section. Accordingly, we generate PDS in $3-6$ keV, $6-15$ keV, $15-25$ keV and $25-60$ keV energy bands, and present the obtained results for AS$2$ in Fig. \ref{fig:PDS_zoom} (left panel). To ascertain the signature of HFQPO and/or harmonic-like features, we fit \texttt{Lorentzian} components at the respective frequencies (Table \ref{table:PDS_parameters}) and compute significance ($\sigma$) and rms amplitudes (in per cent). We find that in $3-6$ keV energy band, a weak feature of HFQPO at $\sim 70$ Hz is present with significance $\sim 1.8\sigma$ and rms $\sim 4.45\%$. However, the possible harmonic at $\sim 150$ Hz is seen to be more prominent with a significance of $\sim 4.3\sigma$ and rms $\sim 6.04\%$. On the contrary, in $6-15$ keV energy band, the evidence of HFQPO is seen to be more prominent with a significance of $\sim 3.2\sigma$ and rms $\sim 6.43\%$, whereas the harmonic-like feature appears to be insignificant. Moreover, the PDS remains featureless in $15-25$ keV and $25-60$ keV energy bands, respectively. In Fig. \ref{fig:PDS_zoom} (right panel), we present the best-fitted energy spectrum (more details in \S\ref{s:spectra}) of the source from combined {\it SXT} and {\it LAXPC} data in $0.7-50$ keV energy band. The model components used in the spectral modeling are indicated at the inset in different colors (\texttt{nthcomp} in red and \texttt{powerlaw} in blue). We observe that the contribution of the Comptonized component (\texttt{nthComp}) remains significant till $\sim 15$ keV, where the evidence of HFQPO and/or harmonic features are observed. This suggests that Comptonized emissions are perhaps attributed to these features.

\section{Spectral Analysis and Results} \label{s:spectra}

We generate energy spectra of the two selected observations (AS$1$ and AS$2$) in the wide-band energy range of $0.7-50$ keV by combining both {\it SXT} and {\it LAXPC} spectra. The {\it SXT} spectra are obtained considering $0.7-7$ keV energy range. We use {\it LAXPC20} spectra in $3-50$ keV energy range which are extracted using \texttt{LaxpcSoftv3.4.4}\footnote{\url{http://www.tifr.res.in/~astrosat\_laxpc/LaxpcSoft.html}} \citep{Antia-etal2017}. We follow the standard procedure to use the background, response and ancilliary files of {\it SXT} and {\it LAXPC} for spectral modeling \cite[see][for details]{Majumder-etal2022}. Each of the combined spectra from {\it SXT} and {\it LAXPC} is modeled using \texttt{XSPEC V12.13.1} in \texttt{HEASOFT V6.32.1} to understand the spectral properties of the source. During spectral fitting, we use the \texttt{gain fit} command to adjust the low energy residuals at $1.8$ keV and $2.2$ keV observed in {\it SXT} spectra. Following \cite{Antia-etal2017}, a systematic error of $2\%$ is incorporated during spectral fitting \cite[]{Leahy-etal2019, Sreehari-etal2020, Majumder-etal2022}.

\begin{figure}
	\begin{center}
		\includegraphics[scale=0.34]{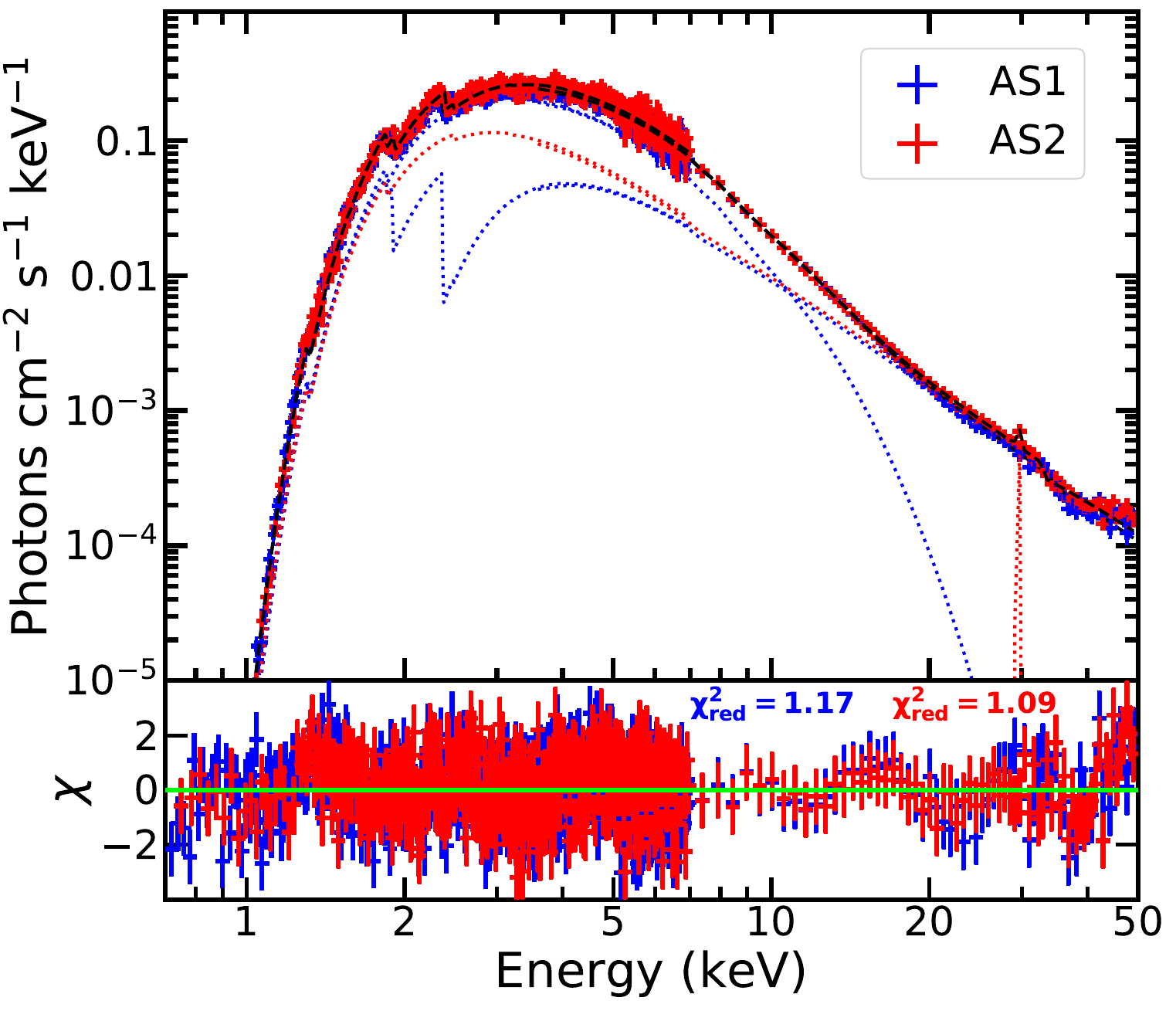}
	\end{center}
	\caption{Best-fitted unfolded wide-band ($0.7-50$ keV) energy spectra of GRS $1915+105$ during Epochs AS$1$ and AS$2$. The spectra are modelled with \texttt{Tbabs$\times$smedge$\times$(nthComp $+$ powerlaw)$\times$constant}. The bottom panel shows the variation of residuals in units of $\sigma$. See text for details.}
	\label{fig:spectrum}
\end{figure}

We adopt the physically motivated model combination comprising of a thermal Comptonization component (\texttt{nthComp}) \cite[]{Zdziarski-etal1996} and a \texttt{powerlaw} component to fit the wide-band spectra in $0.7-50$ keV energy range. The inter-galactic absorption is taken care by including \texttt{Tbabs} \citep{Wilms-etal2000} in the spectral modelling. Following \cite{Majumder-etal2022}, a \texttt{constant} component is included in the spectral fitting to adjust the cross-calibration of the data from {\it SXT} and {\it LAXPC} instruments. During the entire spectral analysis, we keep the column density fixed at $6 \times 10^{22}$ atoms$/$cm$^2$ \citep{Yadav-etal2016,Sreehari-etal2020, Majumder-etal2022}. We find that the above model prescription is inadequate to fit the wide-band spectra of both observations (AS$1$ and AS$2$) with significant residual left near $\sim 9$ keV. Hence, we include a \texttt{smedge} component to adjust the residuals left at $\sim 9$ keV. Accordingly, the final model combination, namely \texttt{Tbabs$\times$smedge$\times$(nthComp $+$ powerlaw)$\times$constant} is found to provide the best description of all the observed spectra with a reasonable reduced chi-square value ($\chi_{\rm red}^{2}=\chi^{2}/dof$) of $1.17$ (AS$1$) and $1.09$ (AS$2$). We note that a \texttt{gaussian} is needed to fit the `bump' present near $\sim 30$ keV in the {\it LAXPC} spectra \citep{Antia-etal21, Aneesha-etal2024} of AS2. In Fig. \ref{fig:spectrum}, we present the best fitted energy spectra of AS1 and AS2 along with residuals using different colors.

The best fitted spectral parameters for observation AS$1$ (AS$2$) are obtained as \texttt{nthComp} photon index $\Gamma_{\rm nth}=1.96_{-0.02}^{+0.05}$ ($1.94_{-0.06}^{+0.08}$), electron temperature $kT_{\rm e}=1.71_{-0.04}^{+0.09}$ ($1.65_{-0.07}^{+0.09}$) keV and power-law photon index $\Gamma_{\rm PL}=2.78_{-0.03}^{+0.04}$ ($2.75_{-0.02}^{+0.03}$), respectively. Note that the seed photon temperature of \texttt{nthcomp} remains unconstrained during the fitting, hence we keep it fixed at $0.2$ keV. We estimate the bolometric flux in $1-100$ keV energy range using the convolution model \texttt{cflux}, keeping the model normalization fixed at the best fitted values. The corresponding bolometric luminosity ($L_{\rm bol}$) is found to be $L_{\rm bol} \sim 0.42~ L_{\rm Edd}$ for both observations, where $L_{\rm Edd}$ refers to the Eddington luminosity. This indicates that the source was in the sub-Eddington accretion regime during Epoch AS$1$ and AS$2$, respectively. Further, following \cite{Zdziarski-etal1996} and using the model fitted spectral parameters, we compute the optical depth as $\sim 13.94$ (AS$1$) and $\sim 14.42$ (AS$2$), and Compton y-parameter as $\sim 2.61$ (AS$1$) and $\sim 2.68$ (AS$2$). This implies the presence of an optically thick Comptonizing corona responsible for the non-thermal emission in the spectra.

\section{Discussion} \label{s:dis}

In this paper, we carry out comprehensive analyses of two GT observations of GRS 1915$+$105 during June 2017 with {\it AstroSat} to examine the variability properties of the source. The obtained findings from the detailed spectro-temporal studies are discussed below.

\subsection{An `Unknown' Variability Class in GRS 1915+105}
 
The light curves during Epoch AS$1$ show the presence of aperiodic `dips' of few tens of seconds duration along with the `flare' like features immediately after the `dip' (Fig. \ref{fig:lcurvCCD}a). The variation of HR also exhibits the signature of the `dip' segments. In the CCD of Epoch AS$1$, we observe a uniform `$C$' shaped distribution with two elongated branches extended towards the higher HR$1$ and HR$2$ domains. Interestingly, the `dip' and `flare' like features in the light curve along with elongated branches in the CCD are not seen in Epoch AS$2$ (Fig. \ref{fig:lcurvCCD}c and Table \ref{table:Obs_details}).

Upon comparing with the existing variability classes observed with {\it AstroSat}, we confirm that this `unknown' class is `softer' \cite[]{Majumder-etal2022} in nature as HR$1 \sim 0.62$ and HR$2 \sim 0.04$. The short duration and aperiodic `dips'/`flares' including the CCD features seem to be different from the $\lambda$, $\omega$ and $\kappa$ classes \cite[]{Belloni-etal2000, Athulya-etal2022}. Further, it is noted that the typical duration of each beat in the canonical $\rho$ class variability of the heartbeat state is about $\sim 60-70$ s and the hardness ratios ($0.5 \lesssim \rm HR1 \lesssim 0.8$ and $0.05 \lesssim \rm HR2 \lesssim 0.2$) indicate towards harder variability properties. Therefore, the `flare' like features (of duration $\sim 40-50$ s) in the light curve of AS$1$ and the corresponding CCD variation appear to be marginally similar to the canonical $\rho$ class. Moreover, in Epoch AS$0$, $16$ days prior to AS$1$, the source exhibits $\rho$ class (Fig. \ref{fig:lcurvCCD}e, Table \ref{table:Obs_details}). During Epoch AS$3$, after $33$ days of AS$2$, the source displays $\kappa$ class variability (Fig. \ref{fig:lcurvCCD}e). We also notice that the combined CCD of $\rho$ and $\kappa$ class observations marginally overlap with that of AS$1$.

Meanwhile, several studies are carried out to explain the origin of $\rho$ class variability in GRS $1915+105$, relying on the classical disc instability model \citep{Lightman-etal1974} for a truncated accretion disc and compact corona configuration \citep{Honma-etal1991, Taam-etal1997, Szuszkiewicz-etal1998, Janiuk-etal2000}. Further, \cite{Massa-etal2013} argued that the limit cycle behaviour, which mostly depends on the accretion rate, is typical for the $\rho$ class variability and explains the hard lags as well. Moreover, \cite{Nandi-etal2000} proposed that the bursts in the light curve of $\rho$ class are closely associated with the so-called `On' state of the source. Most importantly, they found the features of the small-scaled version of a structured $\rho$ class in the light curve of canonical $\kappa$ class variability. This essentially indicates the intrinsic connection between these variability signatures, perhaps produced from the interplay of inner Comptonizing corona and the feedback of disc winds \citep{Nandi-etal2000, Chakrabarti_Nandi_2000}. A phenomenological non-linear mathematical model has also been put forward to describe the complex variability classes of GRS $1915+105$ \citep{Massaro-etal2020a, Massaro-etal2020b}. This model contains an input function $J(t)$ which is used to reproduce light curves of various classes. For example, constant input values \citep{Massaro-etal2020a} reproduce the classes $\chi$, $\delta$ and $\rho$, while several other classes ($\alpha$, $\gamma$, $\lambda$, $\kappa$, $\omega$, $\xi$, and $\theta$) result when rather simple step function or sawtooth modulations of $J(t)$ are considered \citep{Massaro-etal2020b}. We find that the model with a constant input function is unable to reproduce the structured variability signatures consisting of irregular low-count dips and spike-like features as observed in the light curve of the `unknown' variability class ($\tau$). These findings further suggest that the class $\tau$ is quite different from the canonical $\chi$ and $\rho$ classes and possibly could not be explained as a transition between these two classes. However, we cannot exclude the possibility of getting a functional form of $J(t)$, which could model the $\tau$ class variability. It is noteworthy that \cite{Alberti-etal2022} also proposed a formalism based on a stochastic dynamical process to describe the complex variability signatures of GRS $1915+105$. However, applying such models to the present data is beyond the scope of this work and will be addressed elsewhere.

However, the detection of $\delta$ class variability from {\it NuSTAR} observation (Fig. \ref{fig:lcurvCCD}e) within $18$ days of AS$2$ suggests that the observed variability is not just a combined feature of $\rho$ and $\kappa$ classes during $\rho \rightarrow \kappa$ transition, instead it resembles more complex variability signature. Additionally, in Epoch AS$1$ and AS$2$, we observe distinct characteristics, such as low-frequency breaks at $\sim 0.6$ Hz and { $\sim 7$ Hz} in the wide-band ($0.01-500$ Hz) PDS (see Fig. \ref{fig:PDS_QPO}) as compared to the known `softer' variability classes \cite[]{Majumder-etal2022}. Considering all these findings, we infer that the variability patterns seen in the light curves of Epochs AS$1$ and AS$2$ perhaps correspond to an `unknown' variability class compared to the previously reported $15$ known classes of GRS $1915+105$.

\subsection{Possible Detection of Twin Peak HFQPO}

The power spectral analysis indicates the evidence of possible HFQPO features at frequencies $\sim 70$ Hz and $71$ Hz  (Fig. \ref{fig:PDS_QPO}) with rms of $(4.88 \pm 0.97)\%$ and $(4.69 \pm 0.78)\%$ for Epochs AS$1$ and AS$2$, respectively (Table \ref{table:PDS_parameters}). Moreover, in Epoch AS$2$, we find the presence of a harmonic-like feature at $\sim 152$ Hz with rms of $(5.75\pm0.81)\%$ (see Fig. \ref{fig:PDS_QPO} and Table \ref{table:PDS_parameters}). Indeed, for GRS 1915$+$105, the well known HFQPO at $\sim 70$ Hz is already reported with {\it RXTE} \citep{Belloni-etal2013a} and {\it AstroSat} \cite[and reference therein]{Majumder-etal2022}. A broad QPO feature of weak coherence ($\rm Q \sim 2$) at $170$ Hz is also reported during the hard intervals of $\theta$ class with {\it RXTE} \citep{Belloni-etal2006}. 
Interestingly, an additional feature of frequency $27-41$ Hz \citep{Belloni-etal2001, Strohmayer2001b} and twin peak HFQPO feature at $34$ Hz and $68$ Hz \citep{Belloni-etal2013b} are also found from {\it RXTE} observations.
It is noteworthy that twin peak HFQPOs are confirmed in a few other BH-XRBs, such as GRO J1655$-$40 \citep{Remillard-etal1999, Remillard-etal2002}, XTE J1550$-$564 \citep{Remillard-etal2002} and IGR J17091$-$3624 \citep{Altamirano-etal2012}.

The study of energy dependent PDS in Epoch AS$2$ indicates that the possible HFQPO feature at $\sim 70$ Hz is present in $3-15$ keV energy range, whereas the harmonic-like feature only seems to appear in $3-6$ keV. We observe that the strength of the HFQPO feature is more (rms $\sim 6.43\%$) in $6-15$ keV energy band as compared to $3-6$ keV band (rms $\sim 4.45\%$). These findings are coarsely in agreement with the results of \cite{Majumder-etal2022}, except a weak signature of HFQPO present below $\lesssim 6$ keV. 

\subsection{Connection of HFQPOs with Spectral Features}

The wide-band ($0.7-50$ keV) energy spectra of {\it SXT} and {\it LAXPC} are well described by the model consisting of a thermal Comptonization component (\texttt{nthcomp} in \texttt{XSPEC}) and an additional \texttt{powerlaw} component (see \S \ref{s:spectra}). The best-fitted spectra of Epoch AS$1$ (AS$2$) yields an electron temperature $kT_e \sim 1.71 ~(1.65)$ keV, \texttt{nthcomp} photon index $\Gamma_{\rm nth} \sim 1.96 ~(1.94)$, and a steep \texttt{powerlaw} photon index $\Gamma_{\rm PL} \sim 2.78 ~(2.75)$. The source is found to remain in the sub-Eddington regime with bolometric luminosity $L_{\rm bol} \sim 42\%$ $L_{\rm Edd}$ during Epochs AS$1$ and AS$2$ (see \S \ref{s:spectra}). We estimate the optical depth and Compton y-parameter as $\sim 14$ and $\sim 3$ for both observations. This indicates the presence of an optically thick corona responsible for the generation of high energy continuum in the wide-band spectra. Evidently, the requirement of an additional \texttt{powerlaw} component with a steeper index ($\sim 2.8$) corroborates the presence of an extended corona along with the compact `Comptonizing corona' \cite[]{Sreehari-etal2020,Majumder-etal2022}. 

Most importantly, the evidence of possible HFQPO and its harmonic-like feature are found in the energy range ($\lesssim 15$ keV), where Comptonized component dominates in the energy spectra (see Fig. \ref{fig:PDS_zoom}). This possibly implies that only the Comptonized photons within $3-15$ keV energy range participate in such oscillations. Meanwhile, several theoretical studies of accretion dynamics around black holes attempt to explain the HFQPOs observed in various BH-XRBs including GRO J$1655-40$, GRS $1915+105$, XTE J$1550-564$ and H $1743-322$ from the modulation of the boundary of Comptonizing region \citep{Aktar-etal2017, Aktar-etal2018, Dihingia-etal2019}. Moreover, comprehensive spectro-temporal correlation studies of GRS $1915+105$ in the presence of HFQPOs reveal the inherent connection between the origin of these features and the Comptonized emission seen in the energy spectra of four distinct variability classes \citep{Sreehari-etal2020, Majumder-etal2022}. Considering all these and based on the present findings, we speculate that the oscillation of the `Comptonizing corona' seems to play a viable role in exhibiting the possible HFQPOs observed in an `unknown' variability class of GRS $1915+105$.

\section{Conclusions} \label{s:sum}

In this paper, we perform detailed spectro-temporal analyses of GRS $1915+105$ using {\it AstroSat} observations during June, 2017. We observe that the source undergoes a transition from $\rho$ to $\kappa$ class and during such transition, the source displays an { `unknown' variability class ($\tau$)} of GRS $1915+105$, where irregular repetition of low count `dips' along with `flare' like features are observed between the successive steady light curve variations. Note that the source exhibits $\delta$ class variability before transiting to $\kappa$ class, therefore, we ascertain that the `unknown' ($\tau$) type variability is possibly different from the existing $15$ variability classes of GRS 1915$+$105. In addition, we find the evidence of possible HFQPO feature at $\sim 71$ Hz along with a harmonic-like signature at $\sim 152$ Hz. Finally, the wide-band spectral results indicate the presence of a `Compact corona' surrounding the source, which seems to regulate the observed HFQPO features in GRS $1915+105$.

\section{Acknowledgments}
 
Authors thank the anonymous reviewer for constructive comments and useful suggestions that helped to improve the quality of the manuscript. AN thanks Group Head (Space Astronomy Group - SAG), Deputy Director (Payload, Data Management \& Space Astronomy Area - PDMSA) and the Director (U R Rao Satellite Centre - URSC) for encouragement and continuous support to carry out this research. This publication uses the data from the {\it AstroSat} mission of the Indian Space Research Organisation (ISRO), archived at the Indian Space Science Data Centre (ISSDC). This work has used the data from the Soft X-ray Telescope (SXT) developed at TIFR, Mumbai, and the SXT-POC at TIFR is thanked for verifying and releasing the data and providing the necessary software tools. This work has also used the data from the {\it LAXPC} Instruments developed at TIFR, Mumbai, and the LAXPC-POC at TIFR is thanked for verifying and releasing the data. We also thank the {\it AstroSat} Science Support Cell hosted by IUCAA and TIFR for providing the \texttt{LAXPCSOFT} software which we used for {\it LAXPC} data analysis. The authors thank {\it NuSTAR} and {\it RXTE} instrument teams for processing the data and providing the necessary software for analysis.

\section{Data Availability}
{\it AstroSat} data used for this publication are currently available at the Astrobrowse (AstroSat archive) website (\url{https://astrobrowse.issdc.gov.in/astro\_archive/archive}) of the Indian Space Science Data Center (ISSDC). The {\it NuSTAR} and {\it RXTE} data used in this work are archived at the HEASARC data center (\url{https://heasarc.gsfc.nasa.gov/db-perl/W3Browse/w3browse.pl}).

\printendnotes


\begin{thebibliography}{}
\expandafter\ifx\csname natexlab\endcsname\relax\def\natexlab#1{#1}\fi

\bibitem[{{Agrawal}(2006)}]{Agrawal-etal2006}
{Agrawal}, P.~C. 2006, Advances in Space Research, 38, 2989

\bibitem[{{Agrawal} {et~al.}(2017){Agrawal}, {Yadav}, {Antia}, {Dedhia},
  {Shah}, {Chauhan}, {Manchanda}, {Chitnis}, {Gujar}, {Katoch}, {Kurhade},
  {Madhwani}, {Manojkumar}, {Nikam}, {Pandya}, {Parmar}, {Pawar}, {Roy},
  {Paul}, {Pahari}, {Misra}, {Ravichandran}, {Anilkumar}, {Joseph},
  {Navalgund}, {Pandiyan}, {Sarma}, \& {Subbarao}}]{Agrawal-etal2017}
{Agrawal}, P.~C., {Yadav}, J.~S., {Antia}, H.~M., {et~al.} 2017, Journal of
  Astrophysics and Astronomy, 38, 30

\bibitem[{{Agrawal} {et~al.}(2018){Agrawal}, {Nandi}, {Girish}, \&
  {Ramadevi}}]{Agrawal-etal2018}
{Agrawal}, V.~K., {Nandi}, A., {Girish}, V., \& {Ramadevi}, M.~C. 2018, \mnras,
  477, 5437

\bibitem[{{Aktar} {et~al.}(2017){Aktar}, {Das}, {Nandi}, \&
  {Sreehari}}]{Aktar-etal2017}
{Aktar}, R., {Das}, S., {Nandi}, A., \& {Sreehari}, H. 2017, \mnras, 471, 4806

\bibitem[{{Aktar} {et~al.}(2018){Aktar}, {Das}, {Nandi}, \&
  {Sreehari}}]{Aktar-etal2018}
---. 2018, Journal of Astrophysics and Astronomy, 39, 17

\bibitem[{{Alberti} {et~al.}(2022){Alberti}, {Massaro}, {Mineo}, \&
  {Feroci}}]{Alberti-etal2022}
{Alberti}, T., {Massaro}, E., {Mineo}, T., \& {Feroci}, M. 2022, \mnras, 517,
  3568

\bibitem[{{Altamirano} \& {Belloni}(2012)}]{Altamirano-etal2012}
{Altamirano}, D., \& {Belloni}, T. 2012, \apjl, 747, L4

\bibitem[{{Aneesha} {et~al.}(2024){Aneesha}, {Das}, {Katoch}, \&
  {Nandi}}]{Aneesha-etal2024}
{Aneesha}, U., {Das}, S., {Katoch}, T.~B., \& {Nandi}, A. 2024, \mnras, 532,
  4486

\bibitem[{{Antia} {et~al.}(2017){Antia}, {Yadav}, {Agrawal}, {Verdhan Chauhan},
  {Manchanda}, {Chitnis}, {Paul}, {Dedhia}, {Shah}, {Gujar}, {Katoch},
  {Kurhade}, {Madhwani}, {Manojkumar}, {Nikam}, {Pandya}, {Parmar}, {Pawar},
  {Pahari}, {Misra}, {Navalgund}, {Pandiyan}, {Sharma}, \&
  {Subbarao}}]{Antia-etal2017}
{Antia}, H.~M., {Yadav}, J.~S., {Agrawal}, P.~C., {et~al.} 2017, \apjs, 231, 10

\bibitem[{{Antia} {et~al.}(2021){Antia}, {Agrawal}, {Dedhia}, {Katoch},
  {Manchanda}, {Misra}, {Mukerjee}, {Pahari}, {Roy}, {Shah}, \&
  {Yadav}}]{Antia-etal21}
{Antia}, H.~M., {Agrawal}, P.~C., {Dedhia}, D., {et~al.} 2021, Journal of
  Astrophysics and Astronomy, 42, 32

\bibitem[{{Athulya} \& {Nandi}(2023)}]{Athulya-etal2023}
{Athulya}, M.~P., \& {Nandi}, A. 2023, \mnras, 525, 489

\bibitem[{{Athulya} {et~al.}(2022){Athulya}, {Radhika}, {Agrawal},
  {Ravishankar}, {Naik}, {Mandal}, \& {Nandi}}]{Athulya-etal2022}
{Athulya}, M.~P., {Radhika}, D., {Agrawal}, V.~K., {et~al.} 2022, \mnras, 510,
  3019

\bibitem[{{Balakrishnan} {et~al.}(2021){Balakrishnan}, {Miller}, {Reynolds},
  {Kammoun}, {Zoghbi}, \& {Tetarenko}}]{Balakrishnan-etal2021}
{Balakrishnan}, M., {Miller}, J.~M., {Reynolds}, M.~T., {et~al.} 2021, \apj,
  909, 41

\bibitem[{{Belloni} {et~al.}(2000){Belloni}, {Klein-Wolt}, {M{\'e}ndez}, {van
  der Klis}, \& {van Paradijs}}]{Belloni-etal2000}
{Belloni}, T., {Klein-Wolt}, M., {M{\'e}ndez}, M., {van der Klis}, M., \& {van
  Paradijs}, J. 2000, \aap, 355, 271

\bibitem[{{Belloni} {et~al.}(2001){Belloni}, {M{\'e}ndez}, \&
  {S{\'a}nchez-Fern{\'a}ndez}}]{Belloni-etal2001}
{Belloni}, T., {M{\'e}ndez}, M., \& {S{\'a}nchez-Fern{\'a}ndez}, C. 2001, \aap,
  372, 551

\bibitem[{{Belloni} {et~al.}(2006){Belloni}, {Soleri}, {Casella}, {M{\'e}ndez},
  \& {Migliari}}]{Belloni-etal2006}
{Belloni}, T., {Soleri}, P., {Casella}, P., {M{\'e}ndez}, M., \& {Migliari}, S.
  2006, \mnras, 369, 305

\bibitem[{{Belloni} \& {Altamirano}(2013{\natexlab{a}})}]{Belloni-etal2013b}
{Belloni}, T.~M., \& {Altamirano}, D. 2013{\natexlab{a}}, \mnras, 432, 19

\bibitem[{{Belloni} \& {Altamirano}(2013{\natexlab{b}})}]{Belloni-etal2013a}
---. 2013{\natexlab{b}}, \mnras, 432, 10

\bibitem[{{Beri} {et~al.}(2019){Beri}, {Paul}, {Yadav}, {Antia}, {Agrawal},
  {Manchanda}, {Dedhia}, {Chauhan}, {Pahari}, {Misra}, {Katoch}, {Madhwani},
  {Shah}, \& {Varun}}]{Beri-etal2019}
{Beri}, A., {Paul}, B., {Yadav}, J.~S., {et~al.} 2019, \mnras, 482, 4397

\bibitem[{{Casella} {et~al.}(2005){Casella}, {Belloni}, \&
  {Stella}}]{Casella-etal2005}
{Casella}, P., {Belloni}, T., \& {Stella}, L. 2005, \apj, 629, 403

\bibitem[{{Castro-Tirado} {et~al.}(1992){Castro-Tirado}, {Brandt}, \&
  {Lund}}]{Castro-Tirado-etal1992}
{Castro-Tirado}, A.~J., {Brandt}, S., \& {Lund}, N. 1992, IAU, 5590

\bibitem[{{Chakrabarti} \& {Nandi}(2000)}]{Chakrabarti_Nandi_2000}
{Chakrabarti}, S.~K., \& {Nandi}, A. 2000, arXiv e-prints, astro

\bibitem[{{Chakrabarti} {et~al.}(2005){Chakrabarti}, {Nandi}, {Chatterjee},
  {Choudhury}, \& {Chatterjee}}]{Chakrabarti-etal2005}
{Chakrabarti}, S.~K., {Nandi}, A., {Chatterjee}, A.~K., {Choudhury}, A.~K., \&
  {Chatterjee}, U. 2005, \aap, 431, 825

\bibitem[{{Chakrabarti} {et~al.}(2004){Chakrabarti}, {Nandi}, {Choudhury}, \&
  {Chatterjee}}]{Chakrabarti-etal2004}
{Chakrabarti}, S.~K., {Nandi}, A., {Choudhury}, A., \& {Chatterjee}, U. 2004,
  \apj, 607, 406

\bibitem[{{Dihingia} {et~al.}(2019){Dihingia}, {Das}, {Maity}, \& {Nand
  i}}]{Dihingia-etal2019}
{Dihingia}, I.~K., {Das}, S., {Maity}, D., \& {Nand i}, A. 2019, \mnras, 488,
  2412

\bibitem[{{Greiner} {et~al.}(1996){Greiner}, {Morgan}, \&
  {Remillard}}]{Greiner-etal1996}
{Greiner}, J., {Morgan}, E.~H., \& {Remillard}, R.~A. 1996, \apjl, 473, L107

\bibitem[{{Hannikainen} {et~al.}(2005){Hannikainen}, {Rodriguez}, {Vilhu},
  {Hjalmarsdotter}, {Zdziarski}, {Belloni}, {Poutanen}, {Wu}, {Shaw},
  {Beckmann}, {Hunstead}, {Pooley}, {Westergaard}, {Mirabel}, {Hakala},
  {Castro-Tirado}, \& {Durouchoux}}]{Hannikainen-etal2005}
{Hannikainen}, D.~C., {Rodriguez}, J., {Vilhu}, O., {et~al.} 2005, \aap, 435,
  995

\bibitem[{{Harikesh} {et~al.}(2025){Harikesh}, {Majumder}, {Das}, \&
  {Nandi}}]{Harikesh-etal2025}
{Harikesh}, S., {Majumder}, S., {Das}, S., \& {Nandi}, A. 2025, \mnras, 540,
  2965

\bibitem[{{Honma} {et~al.}(1991){Honma}, {Matsumoto}, {Kato}, \&
  {Abramowicz}}]{Honma-etal1991}
{Honma}, F., {Matsumoto}, R., {Kato}, S., \& {Abramowicz}, M.~A. 1991, \pasj,
  43, 261

\bibitem[{{Janiuk} {et~al.}(2000){Janiuk}, {Czerny}, \&
  {Siemiginowska}}]{Janiuk-etal2000}
{Janiuk}, A., {Czerny}, B., \& {Siemiginowska}, A. 2000, \apjl, 542, L33

\bibitem[{Janiuk {et~al.}(2002)Janiuk, Czerny, \&
  Siemiginowska}]{Janiuk-etal2002}
Janiuk, A., Czerny, B., \& Siemiginowska, A. 2002, The Astrophysical Journal,
  576, 908

\bibitem[{{Klein-Wolt} {et~al.}(2002){Klein-Wolt}, {Fender}, {Pooley},
  {Belloni}, {Migliari}, {Morgan}, \& {van der Klis}}]{Klein-wolt-etal2002}
{Klein-Wolt}, M., {Fender}, R.~P., {Pooley}, G.~G., {et~al.} 2002, \mnras, 331,
  745

\bibitem[{{Leahy} \& {Chen}(2019)}]{Leahy-etal2019}
{Leahy}, D.~A., \& {Chen}, Y. 2019, \apj, 871, 152

\bibitem[{{Leahy} {et~al.}(1983){Leahy}, {Darbro}, {Elsner}, {Weisskopf},
  {Sutherland}, {Kahn}, \& {Grindlay}}]{Leahy-etal1983}
{Leahy}, D.~A., {Darbro}, W., {Elsner}, R.~F., {et~al.} 1983, \apj, 266, 160

\bibitem[{{Lightman} \& {Eardley}(1974)}]{Lightman-etal1974}
{Lightman}, A.~P., \& {Eardley}, D.~M. 1974, \apjl, 187, L1

\bibitem[{{Lotti} {et~al.}(2016){Lotti}, {Natalucci}, {Mori}, {Baganoff},
  {Boggs}, {Christensen}, {Craig}, {Hailey}, {Harrison}, {Hong}, {Krivonos},
  {Rahoui}, {Stern}, {Tomsick}, {Zhang}, \& {Zhang}}]{Lotti-etal2016}
{Lotti}, S., {Natalucci}, L., {Mori}, K., {et~al.} 2016, \apj, 822, 57

\bibitem[{{Majumder} {et~al.}(2024){Majumder}, {Dutta}, \&
  {Nandi}}]{Majumder-etal2024}
{Majumder}, P., {Dutta}, B.~G., \& {Nandi}, A. 2024, \mnras, 527, 4739

\bibitem[{{Majumder} {et~al.}(2025){Majumder}, {Dutta}, \&
  {Nandi}}]{Majumder-etal2025}
---. 2025, \mnras, 540, 37

\bibitem[{{Majumder} {et~al.}(2022){Majumder}, {Sreehari}, {Aftab}, {Katoch},
  {Das}, \& {Nandi}}]{Majumder-etal2022}
{Majumder}, S., {Sreehari}, H., {Aftab}, N., {et~al.} 2022, \mnras, 512, 2508

\bibitem[{{Massa} {et~al.}(2013){Massa}, {Massaro}, {Mineo}, {D'A{\`\i}},
  {Feroci}, {Casella}, \& {Belloni}}]{Massa-etal2013}
{Massa}, F., {Massaro}, E., {Mineo}, T., {et~al.} 2013, \aap, 556, A84

\bibitem[{{Massaro} {et~al.}(2020{\natexlab{a}}){Massaro}, {Capitanio},
  {Feroci}, {Mineo}, {Ardito}, \& {Ricciardi}}]{Massaro-etal2020a}
{Massaro}, E., {Capitanio}, F., {Feroci}, M., {et~al.} 2020{\natexlab{a}},
  \mnras, 495, 1110

\bibitem[{{Massaro} {et~al.}(2020{\natexlab{b}}){Massaro}, {Capitanio},
  {Feroci}, {Mineo}, {Ardito}, \& {Ricciardi}}]{Massaro-etal2020b}
---. 2020{\natexlab{b}}, \mnras, 496, 1697

\bibitem[{{M{\'e}ndez} {et~al.}(2013){M{\'e}ndez}, {Altamirano}, {Belloni}, \&
  {Sanna}}]{Mendez-etal2013}
{M{\'e}ndez}, M., {Altamirano}, D., {Belloni}, T., \& {Sanna}, A. 2013, \mnras,
  435, 2132

\bibitem[{{Morgan} {et~al.}(1999){Morgan}, {Remillard}, {Muno}, \&
  {Kitzgibbons}}]{Morgan-etal1999}
{Morgan}, E.~H., {Remillard}, R., {Muno}, M., \& {Kitzgibbons}, K. 1999, in
  AAS/High Energy Astrophysics Division, Vol.~4, AAS/High Energy Astrophysics
  Division \#4, 28.11

\bibitem[{{Motta} {et~al.}(2021){Motta}, {Kajava}, {Giustini}, {Williams}, {Del
  Santo}, {Fender}, {Green}, {Heywood}, {Rhodes}, {Segreto}, {Sivakoff}, \&
  {Woudt}}]{Motta-etal2021}
{Motta}, S.~E., {Kajava}, J.~J.~E., {Giustini}, M., {et~al.} 2021, \mnras, 503,
  152

\bibitem[{{Muno} {et~al.}(1999){Muno}, {Morgan}, \&
  {Remillard}}]{Muno-etal1999}
{Muno}, M.~P., {Morgan}, E.~H., \& {Remillard}, R.~A. 1999, arXiv e-prints,
  astro

\bibitem[{{Naik} {et~al.}(2002){Naik}, {Agrawal}, {Rao}, \&
  {Paul}}]{Naik-etal2002}
{Naik}, S., {Agrawal}, P.~C., {Rao}, A.~R., \& {Paul}, B. 2002, \mnras, 330,
  487

\bibitem[{{Nandi} {et~al.}(2012){Nandi}, {Debnath}, {Mandal}, \&
  {Chakrabarti}}]{Nandi-etal2012}
{Nandi}, A., {Debnath}, D., {Mandal}, S., \& {Chakrabarti}, S.~K. 2012, \aap,
  542, A56

\bibitem[{{Nandi} {et~al.}(2000){Nandi}, {Manickam}, \&
  {Chakrabarti}}]{Nandi-etal2000}
{Nandi}, A., {Manickam}, S.~G., \& {Chakrabarti}, S.~K. 2000, arXiv e-prints,
  astro

\bibitem[{{Nandi} {et~al.}(2001){Nandi}, {Manickam}, {Rao}, \&
  {Chakrabarti}}]{Nandi-etal2001}
{Nandi}, A., {Manickam}, S.~G., {Rao}, A.~R., \& {Chakrabarti}, S.~K. 2001,
  \mnras, 324, 267

\bibitem[{{Nathan} {et~al.}(2022){Nathan}, {Ingram}, {Homan}, {Huppenkothen},
  {Uttley}, {van der Klis}, {Motta}, {Altamirano}, \&
  {Middleton}}]{Nathan-etal2022}
{Nathan}, E., {Ingram}, A., {Homan}, J., {et~al.} 2022, \mnras, 511, 255

\bibitem[{{Neilsen} {et~al.}(2011){Neilsen}, {Remillard}, \&
  {Lee}}]{Neilsen-etal2011}
{Neilsen}, J., {Remillard}, R.~A., \& {Lee}, J.~C. 2011, \apj, 737, 69

\bibitem[{{Pahari} \& {Pal}(2009)}]{Pahari-etal2009}
{Pahari}, M., \& {Pal}, S. 2009, arXiv e-prints, arXiv:0906.4611

\bibitem[{{Parrinello} {et~al.}(2023){Parrinello}, {Neilsen}, {Homan},
  {Steiner}, {Uttley}, \& {Cackett}}]{Parrinello-etal2023}
{Parrinello}, K., {Neilsen}, J., {Homan}, J., {et~al.} 2023, in American
  Astronomical Society Meeting Abstracts, Vol.~55, American Astronomical
  Society Meeting Abstracts, 469.01

\bibitem[{{Paul} {et~al.}(1997){Paul}, {Agrawal}, {Rao}, {Vahia}, {Yadav},
  {Marar}, {Seetha}, \& {Kasturirangan}}]{Paul-etal1997}
{Paul}, B., {Agrawal}, P.~C., {Rao}, A.~R., {et~al.} 1997, \aap, 320, L37

\bibitem[{{Paul} {et~al.}(1998{\natexlab{a}}){Paul}, {Agrawal}, {Rao}, {Vahia},
  {Yadav}, {Marar}, {Seetha}, \& {Kasturirangan}}]{Paul-etal1998a}
{Paul}, B., {Agrawal}, P.~C., {Rao}, A.~R., {et~al.} 1998{\natexlab{a}}, in The
  Hot Universe, ed. K.~{Koyama}, S.~{Kitamoto}, \& M.~{Itoh}, Vol. 188, 394

\bibitem[{{Paul} {et~al.}(1998{\natexlab{b}}){Paul}, {Agrawal}, {Rao}, {Vahia},
  {Yadav}, {Marar}, {Seetha}, \& {Kasturirangan}}]{Paul-etal1998c}
---. 1998{\natexlab{b}}, \aaps, 128, 145

\bibitem[{{Paul} {et~al.}(1998{\natexlab{c}}){Paul}, {Agrawal}, {Rao}, {Vahia},
  {Yadav}, {Seetha}, \& {Kasturirangan}}]{Paul-etal1998b}
---. 1998{\natexlab{c}}, \apjl, 492, L63

\bibitem[{{Rawat} {et~al.}(2019){Rawat}, {Pahari}, {Yadav}, {Jain}, {Misra},
  {Bagri}, {Katoch}, {Agrawal}, \& {Manchanda}}]{Rawat-etal2019}
{Rawat}, D., {Pahari}, M., {Yadav}, J.~S., {et~al.} 2019, \apj, 870, 4

\bibitem[{{Remillard} {et~al.}(2006){Remillard}, {McClintock}, {Orosz}, \&
  {Levine}}]{Remillard-etal_2006}
{Remillard}, R.~A., {McClintock}, J.~E., {Orosz}, J.~A., \& {Levine}, A.~M.
  2006, \apj, 637, 1002

\bibitem[{{Remillard} {et~al.}(1999){Remillard}, {Morgan}, {McClintock},
  {Bailyn}, \& {Orosz}}]{Remillard-etal1999}
{Remillard}, R.~A., {Morgan}, E.~H., {McClintock}, J.~E., {Bailyn}, C.~D., \&
  {Orosz}, J.~A. 1999, \apj, 522, 397

\bibitem[{{Remillard} {et~al.}(2002){Remillard}, {Muno}, {McClintock}, \&
  {Orosz}}]{Remillard-etal2002}
{Remillard}, R.~A., {Muno}, M.~P., {McClintock}, J.~E., \& {Orosz}, J.~A. 2002,
  \apj, 580, 1030

\bibitem[{{Rodriguez} {et~al.}(2004){Rodriguez}, {Corbel}, {Hannikainen},
  {Belloni}, {Paizis}, \& {Vilhu}}]{Rodriguez-etal2004}
{Rodriguez}, J., {Corbel}, S., {Hannikainen}, D.~C., {et~al.} 2004, \apj, 615,
  416

\bibitem[{{Singh} {et~al.}(2014){Singh}, {Tandon}, {Agrawal}, {Antia},
  {Manchanda}, {Yadav}, {Seetha}, {Ramadevi}, {Rao}, {Bhattacharya}, {Paul},
  {Sreekumar}, {Bhattacharyya}, {Stewart}, {Hutchings}, {Annapurni}, {Ghosh},
  {Murthy}, {Pati}, {Rao}, {Stalin}, {Girish}, {Sankarasubramanian},
  {Vadawale}, {Bhalerao}, {Dewangan}, {Dedhia}, {Hingar}, {Katoch}, {Kothare},
  {Mirza}, {Mukerjee}, {Shah}, {Shah}, {Mohan}, {Sangal}, {Nagabhusana},
  {Sriram}, {Malkar}, {Sreekumar}, {Abbey}, {Hansford}, {Beardmore}, {Sharma},
  {Murthy}, {Kulkarni}, {Meena}, {Babu}, \& {Postma}}]{Singh-etal2014}
{Singh}, K.~P., {Tandon}, S.~N., {Agrawal}, P.~C., {et~al.} 2014, in Society of
  Photo-Optical Instrumentation Engineers (SPIE) Conference Series, Vol. 9144,
  Space Telescopes and Instrumentation 2014: Ultraviolet to Gamma Ray, ed.
  T.~{Takahashi}, J.-W.~A. {den Herder}, \& M.~{Bautz}, 91441S

\bibitem[{{Singh} {et~al.}(2016){Singh}, {Stewart}, {Chandra}, {Mukerjee},
  {Kotak}, {Beardmore}, {Chitnis}, {Dewangan}, {Bhattacharyya}, {Mirza},
  {Kamble}, {Navalkar}, {Shah}, {Vishwakarma}, \& {Koyande}}]{Singh-etal2016}
{Singh}, K.~P., {Stewart}, G.~C., {Chandra}, S., {et~al.} 2016, in Society of
  Photo-Optical Instrumentation Engineers (SPIE) Conference Series, Vol. 9905,
  Space Telescopes and Instrumentation 2016: Ultraviolet to Gamma Ray, ed.
  J.-W.~A. {den Herder}, T.~{Takahashi}, \& M.~{Bautz}, 99051E

\bibitem[{{Singh} {et~al.}(2017){Singh}, {Stewart}, {Westergaard},
  {Bhattacharayya}, {Chandra}, {Chitnis}, {Dewangan}, {Kothare}, {Mirza},
  {Mukerjee}, {Navalkar}, {Shah}, {Abbey}, {Beardmore}, {Kotak}, {Kamble},
  {Vishwakarama}, {Pathare}, {Risbud}, {Koyande}, {Stevenson}, {Bicknell},
  {Crawford}, {Hansford}, {Peters}, {Sykes}, {Agarwal}, {Sebastian},
  {Rajarajan}, {Nagesh}, {Narendra}, {Ramesh}, {Rai}, {Navalgund}, {Sarma},
  {Pandiyan}, {Subbarao}, {Gupta}, {Thakkar}, {Singh}, \&
  {Bajpai}}]{Singh-etal2017}
{Singh}, K.~P., {Stewart}, G.~C., {Westergaard}, N.~J., {et~al.} 2017, Journal
  of Astrophysics and Astronomy, 38, 29

\bibitem[{{Sreehari} {et~al.}(2020){Sreehari}, {Nandi}, {Das}, {Agrawal},
  {Mandal}, {Ramadevi}, \& {Katoch}}]{Sreehari-etal2020}
{Sreehari}, H., {Nandi}, A., {Das}, S., {et~al.} 2020, \mnras, 499, 5891

\bibitem[{{Sreehari} {et~al.}(2019){Sreehari}, {Ravishankar}, {Iyer},
  {Agrawal}, {Katoch}, {Mandal}, \& {Nand i}}]{Sreehari-etal2019}
{Sreehari}, H., {Ravishankar}, B.~T., {Iyer}, N., {et~al.} 2019, \mnras, 487,
  928

\bibitem[{{Strohmayer}(2001{\natexlab{a}})}]{Strohmayer2001a}
{Strohmayer}, T.~E. 2001{\natexlab{a}}, \apjl, 552, L49

\bibitem[{{Strohmayer}(2001{\natexlab{b}})}]{Strohmayer2001b}
---. 2001{\natexlab{b}}, \apjl, 554, L169

\bibitem[{{Szuszkiewicz} \& {Miller}(1998)}]{Szuszkiewicz-etal1998}
{Szuszkiewicz}, E., \& {Miller}, J.~C. 1998, \mnras, 298, 888

\bibitem[{{Taam} {et~al.}(1997){Taam}, {Chen}, \& {Swank}}]{Taam-etal1997}
{Taam}, R.~E., {Chen}, X., \& {Swank}, J.~H. 1997, \apjl, 485, L83

\bibitem[{{Uttley} {et~al.}(2014){Uttley}, {Cackett}, {Fabian}, {Kara}, \&
  {Wilkins}}]{Uttley-etal2014}
{Uttley}, P., {Cackett}, E.~M., {Fabian}, A.~C., {Kara}, E., \& {Wilkins},
  D.~R. 2014, \aapr, 22, 72

\bibitem[{{van der Klis}(1988)}]{VanderKlis1989}
{van der Klis}, M. 1988, in NATO Advanced Science Institutes (ASI) Series C,
  Vol. 262, NATO Advanced Science Institutes (ASI) Series C, ed.
  H.~{{\"O}gelman} \& E.~P.~J. {van den Heuvel} (Kluwer Academic Publishers,
  Dordrecht), 27

\bibitem[{{Vaughan} {et~al.}(2003){Vaughan}, {Edelson}, {Warwick}, \&
  {Uttley}}]{Vaughan-etal2003}
{Vaughan}, S., {Edelson}, R., {Warwick}, R.~S., \& {Uttley}, P. 2003, \mnras,
  345, 1271

\bibitem[{{Vilhu} \& {Nevalainen}(1998)}]{Vilhu-etal1998}
{Vilhu}, O., \& {Nevalainen}, J. 1998, \apjl, 508, L85

\bibitem[{{Weng} {et~al.}(2018){Weng}, {Wang}, {Cai}, {Yuan}, \&
  {Gu}}]{Weng-etal2018}
{Weng}, S.-S., {Wang}, T.-T., {Cai}, J.-P., {Yuan}, Q.-R., \& {Gu}, W.-M. 2018,
  \apj, 865, 19

\bibitem[{{Wilms} {et~al.}(2000){Wilms}, {Allen}, \& {McCray}}]{Wilms-etal2000}
{Wilms}, J., {Allen}, A., \& {McCray}, R. 2000, \apj, 542, 914

\bibitem[{{Yadav} {et~al.}(1999){Yadav}, {Rao}, {Agrawal}, {Paul}, {Seetha}, \&
  {Kasturirangan}}]{Yadav-etal1999}
{Yadav}, J.~S., {Rao}, A.~R., {Agrawal}, P.~C., {et~al.} 1999, \apj, 517, 935

\bibitem[{{Yadav} {et~al.}(2016){Yadav}, {Agrawal}, {Antia}, {Chauhan},
  {Dedhia}, {Katoch}, {Madhwani}, {Manchanda}, {Misra}, {Pahari}, {Paul}, \&
  {Shah}}]{Yadav-etal2016}
{Yadav}, J.~S., {Agrawal}, P.~C., {Antia}, H.~M., {et~al.} 2016, in Society of
  Photo-Optical Instrumentation Engineers (SPIE) Conference Series, Vol. 9905,
  Space Telescopes and Instrumentation 2016: Ultraviolet to Gamma Ray, ed.
  J.-W.~A. {den Herder}, T.~{Takahashi}, \& M.~{Bautz}, 99051D

\bibitem[{{Zdziarski} {et~al.}(1996){Zdziarski}, {Johnson}, \&
  {Magdziarz}}]{Zdziarski-etal1996}
{Zdziarski}, A.~A., {Johnson}, W.~N., \& {Magdziarz}, P. 1996, \mnras, 283, 193

\bibitem[{{Zhang} {et~al.}(1995){Zhang}, {Jahoda}, {Swank}, {Morgan}, \&
  {Giles}}]{Zhang-etal1995}
{Zhang}, W., {Jahoda}, K., {Swank}, J.~H., {Morgan}, E.~H., \& {Giles}, A.~B.
  1995, \apj, 449, 930

\bibitem[{{Zhang} {et~al.}(2022){Zhang}, {M{\'e}ndez}, {Garc{\'\i}a},
  {Karpouzas}, {Zhang}, {Liu}, {Belloni}, \& {Altamirano}}]{Zhang-etal2022}
{Zhang}, Y., {M{\'e}ndez}, M., {Garc{\'\i}a}, F., {et~al.} 2022, \mnras, 514,
  2891

\end{thebibliography}
\end{document}